\documentclass[aps,prd,onecolumn,eqsecnum,amsmath,nofootinbib,preprintnumbers]{revtex4}%

\usepackage{color,graphicx,float,subfigure}
\usepackage{amsfonts,amssymb,theorem,mathrsfs,times}
\usepackage{bm}
\usepackage{mathtools}
\usepackage{amsfonts,amssymb,theorem,mathrsfs}
\usepackage{dsfont}
\usepackage{setspace}
\usepackage{graphicx}
\usepackage{ulem}
\usepackage[colorlinks,linkcolor=blue,anchorcolor=blue,citecolor=blue]{hyperref}
{\theorembodyfont{\upshape}
}
{\theorembodyfont{\upshape}
}
{\theorembodyfont{\upshape}
}
{\theorembodyfont{\upshape}
}
{\theorembodyfont{\upshape}
}
{\theorembodyfont{\upshape}
}

\newcommand{\dalm}{\kern1pt\vbox{\hrule height 0.9pt\hbox{\vrule width
0.9pt\hskip 2.5pt\vbox{\vskip 5.5pt}\hskip 3pt\vrule width
0.3pt}\hrule height 0.3pt}\kern1pt}

\begin{document}
\preprint{\hfill {\small{ICTS-USTC/PCFT-24-05}}}
\title{The appearance of de Sitter black holes and strong cosmic censorship
}

%

\author{ Li-Ming Cao$^{a\, ,b}$\footnote{e-mail
address: caolm@ustc.edu.cn}}

\author{ Long-Yue Li$^b$\footnote{e-mail
address: lily26@mail.ustc.edu.cn}}

\author{ Xia-Yuan Liu$^b$ \footnote{e-mail
address: liuxiayuan@mail.ustc.edu.cn }}

\author{ Yu-Sen Zhou$^b$\footnote{e-mail
address: zhou\_ys@mail.ustc.edu.cn}}

\affiliation{$^a$Peng Huanwu Center for Fundamental Theory, Hefei, Anhui 230026, China}

\affiliation{${}^b$
Interdisciplinary Center for Theoretical Study and Department of Modern Physics,\\
University of Science and Technology of China, Hefei, Anhui 230026,
China}


\date{\today}


\begin{abstract}
We study the optical appearance of Schwarzschild-de Sitter and Reissner-Nordstr\"{o}m-de Sitter black holes viewed by  distant observers inside cosmological horizons.
Unlike their asymptotically flat counterparts, due to the  positive cosmological constant,
there are outermost stable circular orbits in the spacetimes, resulting in significant outer edges in the images.
Besides, when the Reissner-Nordstr\"{o}m-de Sitter black hole has a stable Cauchy horizon, the photons from the preceding
companion universe can be received by the observer in our universe.
These rays create a multi-ring structure in the image.
Since the stable Cauchy horizon violates the strong cosmic censorship conjecture,
this novel image shed some light on the test of the conjecture by astronomical observations.
\end{abstract}


\maketitle


\section{Introduction}

Despite its incredible success, general relativity (GR) indicates its own breakdown at singularities.
This was demonstrated by Hawking and Penrose \cite{Penrose:1964wq,Hawking:1970zqf}, who showed that within the framework of GR, gravitational collapse inevitably leads to singularities.
These singularities mark the catastrophic edge where determinism breaks down in our current understanding of physics.
To remedy this issue, two conjectures were proposed, the weak cosmic censorship conjecture (WCCC) \cite{Penrose:1969pc} and the strong cosmic censorship conjecture (SCCC)~\cite{Penrose1974-1}.
The WCCC primarily concerns on the visibility of singularities by far away observers, postulating that singularities are always concealed behind a horizon.
For instance, a Kerr-Newman (KN) black hole can not be over charged or over  spun to become a naked KN singularity by absorbing matter \cite{Wald:1974hkz,Sorce:2017dst}.
On the other hand, the SCCC is devoted to rescue the determinism of GR.
It states that a physical spacetime is always globally hyperbolic \cite{Penrose1974-1,Penrose:1978,Penrose:1979azm}.
However, The extension of spacetime beyond the Cauchy horizon processed in Reissner-Nordstr\"{o}m (RN) black hole seems to violate the SCCC.
Fortunately, the Cauchy horizon of the RN black hole has been shown to be unstable~\cite{Simpson:1973ua,Mcnamara:1977,Mcnamara:1978,1982RSPSA.384..301C, Hiscock1991,Poisson:1990eh,Ori:1991zz}.
Several methods have been proposed to examine the validity of the SCCC within a given spacetime.
For example, the perturbation of RN black hole grows unbounded near the Cauchy horizon~\cite{Simpson:1973ua,Mcnamara:1977,Mcnamara:1978,1982RSPSA.384..301C}.
And the method involved the backreation  by considering ingoing null flux~\cite{Hiscock1991}.
Furthermore, Poisson and Israel \cite{Poisson:1990eh} found that the mass inflation, i.e., the divergence of the Hawking mass or renormalized Hawking mass, can be triggered by the presence of an outgoing null flux.
Therefore, the Cauchy horizon of RN black hole will convert into a null singularity after a small perturbation, which prevents the observer to pass through it.
The Cauchy horizons of many other black holes, whose Penrose diagrams are consist of repeated universes as RN black hole, are also unstable \cite{Cao:2023aco,Brown:2011tv,Iofa:2022dnc,Carballo-Rubio:2021bpr}.

It was commonly believed that the SCCC holds true and may only be violated in some peculiar modified gravity or when extraordinary matter is present.
However, even in the context of GR, which is the most well-established theory of gravity, counterexamples exist.
One such counterexample is the RN-de Sitter (dS) black hole, which is distinguished by a single additional parameter compared to the RN black hole.
It is one of the most typical solutions with a non-vanishing cosmological constant.
Under certain parameters, the RN-dS black hole can exhibit a stable Cauchy horizon \cite{Cardoso:2017soq,Costa:2015pwa,Costa:2016afl,Brady:1992cz,Mellor:1989ac}.
Consequently, the Penrose diagram of the RN-dS black hole can be extended to be consisted of countless repeated identical universes. Therefore, observers could travel through a black-white hole bridge and then enter another separate universe, which violates the SCCC.
More precisely, the Cauchy horizon remains stable when the surface gravity of the Cauchy horizon is less than that of the cosmological horizon \cite{Brady:1992,Cai:1995ux}.
The mass inflation in more sophisticate  Einstein-Maxwell-scalar field model has been thoroughly studied in the trilogy~\cite{Costa:2014yha,Costa:2014zha,Costa:2014aia,Rossetti:2023mrx}, in particular, a detailed range of parameters for mass inflation has been achieved.

It is widely believed in astronomy that massive stars will eventually collapse to form rotating black holes, i.e., Kerr black holes.
There are many similarities between RN black holes and Kerr one.
For example, they both have singularities and Cauchy horizons.
The Penrose diagram of the equatorial plane of Kerr black hole is similar to that of RN black hole.
This implies that the global structure of Kerr black hole is similar to that of RN one to some extent.
Just as the cosmological constant stabilizes the Cauchy horizon of the RN black hole, similar scenarios occur for the Kerr black hole. In other words,
the Cauchy horizon of the Kerr black hole is also unstable, which respects SCCC \cite{Ori:1992zz}.
However, when the cosmological constant is taken into account, the Cauchy horizon of the Kerr-dS black hole can be stable for certain parameters, which violates SCCC \cite{Chambers:1997ef}. On the other hand,
a positive cosmological constant can describe the inflation of the early universe and plays a crucial role in explaining the universe's accelerating expansion.
Multiple observations now also favor a positive cosmological constant.
Therefore, the Kerr-dS black hole has garnered considerable interest.
In short, the similarities mentioned above even appear when they are immersed in de Sitter space.
Based on the similarities, the study of SCCC in RN-dS black hole has reference significance for the study of Kerr-dS one and could potentially serve as a natural experimental site for testing the SCCC.

Several years ago, the images of the supermassive objects at the center of M87 \cite{EventHorizonTelescope:2019dse} and Milky Way galaxies \cite{EventHorizonTelescope:2022wkp} was captured by the Event Horizon Telescope (EHT).
This provides us a  novel method to detect the compact objects and the black holes.
In these black hole pictures, a bright ring surrounds a dark shadow, formed by light emitted directly from the accretion disk.
According to calculations for the null geodesics, light rays emitted from the vicinity of the photon sphere, where the effective potential reaches its maximum, can circulate around the black hole numerous times before being received by the observer.
A more general definition of the photon sphere (surface) can be found in \cite{Claudel:2000yi}.
However, despite anticipating them, this type of light remained undetected on these images.
Utilizing ray-tracing method \cite{Gralla:2019xty}, we can trace light rays backwards from the observer's image plane.
The trajectory of light rays intersects with the accretion disk, subsequently converging to a bright ring in the image plane.
These rings is called a ``lensed ring" for two intersections, and ``photon ring" for more intersections.
However, the Lyapunov exponent being positive indicates that the photon sphere is chaotic. This means that the separation between two initially close trajectories will diverge exponentially over time.
Consequently, the photon ring is so close to the lensed ring that it can not be distinguished apart, and is much narrower that its contribution to the overall intensity is negligible.
The lensed rings are also covered by the directly emitted light due to the current limitations of astronomical observation precision.
As the impact parameter reaches its critical limit, the rays rotate at the photon sphere for infinite times.
Since the accretion disk is distributed in a  limited range, the rays emitted from the edge of the source will form a sharp edge on the image.
The position of the lensed ring or photon ring reveals the information about the background geometry, while the total appearance of a compact object is heavily influenced by the position and profile of the light source.

Recently the images of many compact objects were studied by the ray-tracing method proposed in~\cite{Gralla:2019xty}, such as Kazakov-Solodukhin black hole, the regular black hole with de Sitter core, the quantum-corrected black hole, the wormholes and so on~
\cite{Peng:2020wun,DeMartino:2023ovj,Wang:2023rjl,Zhang:2023okw,Huang:2023yqd,Luo:2023wru,Peng:2021osd,Guo:2022iiy}.
All of these works have exhibited the reliability of the way to get the optical appearance of the black holes or other compact objects. Therefore, it is valuable to apply the method to study the image of the black hole when the SCCC is destroyed, and find some specific and indicative features which might provide some enlightenment on the astronomical observation.
Actually,
in \cite{Cao:2023par}, we have studied the image of a regular black hole with a stable Cauchy horizon, which is inconsistent with SCCC.
The rays from the preceding companion universe can be received by the observer in our universe due to the stable Cauchy horizon.
This produces many new rings inside the shadow area in the image.
This novel multi-rings structure may be detected astronomically.
The fly in the ointment is that a precise physical process that leads to the formation of such black hole still remains a mystery.
On the other hand, it is clear that the RN-dS black hole with a stable Cauchy horizon has a reasonable explanation in the framework of Einstein gravity theory. It is natural to exploit its image in the same way. However, unlike the asymptotically flat black holes, some difficulties arise when the positive cosmological constant is presented.
Firstly, for the RN-dS black hole, the observers have to be located inside the cosmological horizon rather than the null infinity.
But the image is influenced greatly by the location of the observer while there is no specific location is given privileged status. The image has an outer edge if the observer is close to the black hole.
And the image formed by the rays emitted directly is more narrow compared with that of Schwarzschild
one.
Secondly, unlike the asymptotically flat black holes, there are  outermost stable circular orbits (OSCO) in the spacetimes of the  dS or charged dS black hole \cite{Berry:2020ntz,Boonserm:2019nqq,Song:2021ziq}.
A physically reasonable thin accretion disk is distributed between innermost stable circular orbit (ISCO) and OSCO.
And this leads to a significant outer edge in the image.
Thirdly, during the propagation of light, the redshift or blueshift factor in RN-dS black hole is quite different from the one in asymptotically flat case, for example, it tends to infinity near the cosmological horizon.
A very large intensity of light will be received by the observers near the cosmological horizon.
All of these will be discussed in detail in section \ref{S3}.

As the regular black hole with a stable Cauchy horizon, the stable Cauchy horizon in RN-dS black hole also gives rise to the multi-rings structure.
Hence it may be a potential method to test SCCC in astronomical observations.
However, the multi-rings structure also occurs in the compact object without horizons and the wormholes \cite{Olmo:2021piq,Guerrero:2022qkh,Guerrero:2022msp,Olmo:2023lil}.
And the similar multi images structure can be found in the image of stars caused by gravitational lensing of a massive object \cite{Tsukamoto:2021caq,Tsukamoto:2022vkt}.
The difference between these objects and the black hole violating SCCC is whether they have horizons.
In \cite{Morales-Herrera:2024zbr}, a general relativistic formalism is proposed to get the parameter of the RN black hole (such as the mass and the charge) from some directly observable quantities (such as the total frequency shift, aperture angle of the telescope, and redshift rapidity).
And a method to get the parameter of the Kerr-dS black hole is proposed in \cite{Momennia:2023lau}.
Maybe we can distinguish the compact object, wormhole and the RN-dS black hole with the assistance of multiple methods, especially the experiment involving a significant redshift in the presence of an event horizon.

This paper is organized as follows.
In section \ref{S2}, we introduce the ISCO and OSCO of the RN-dS black hole and get the range when there is a stable Cauchy horizon, an ISCO and OSCO in RN-dS spacetime.
In section \ref{S3}, we draw up the image of the Schwarzschild-dS black hole and analyse the effect of observer distance on the image.
Then the image of the RN-dS black hole is investigated in \ref{S4}.
Finally, we give conclusions and discussions in section \ref{S5}.

\section{The ISCO and OSCO in RN-dS black hole}\label{S2}
The existence of an ISCO for massive particles in the Schwarzschild black hole at $r = 6m$ is a well-known fact.
However, in addition to the ISCO, there is an OSCO in the spacetime of the Schwarzschild-dS black hole or more general RN-dS black hole.
Physically acceptable substances that can emit light are generally massive.
As a result, we adapt the model of the accretion disk, which is regarded as the source of light emission, to consist of the stable orbits of these massive substances.
Consequently, the accretion disk in an RN-dS black hole resembles a band distributed between the ISCO and OSCO.
The accretion disk of a Kerr-dS spacetime is studied in \cite{Stuchlik:2020rls}.
In this section, we will give a brief introduction to the ISCO and OSCO of the RN-dS black hole.

In static coordinates, the metric of RN-dS black hole can be written as
\begin{eqnarray}
\mathrm{d}s^2=-f(r)\mathrm{d}t^2+\frac{1}{f(r)}\mathrm{d}r^2+r^2\mathrm{d}\theta^2+r^2\sin^2\theta \mathrm{d}\phi^2  \,,
\end{eqnarray}
where
\begin{eqnarray}
f(r)=1-\frac{2M}{r}+\frac{Q^2}{r^2}-\frac{\Lambda}{3}r^2    \,.
\end{eqnarray}
Here, $M$ is mass parameter, $Q$ is electric charge parameter, and $\Lambda>0$ is the cosmological constant. We are interested in case with three horizons, i.e., the Cauchy horizon, the event horizon, and the cosmological horizon. The radii of these horizons are denoted as $r_-$, $r_+$, and $r_c$, respectively, which are three real roots of $f(r)=0$.
 With the help of the Killing vector fields $\partial/\partial t$ and $\partial/\partial \phi$, we can get the conserved quantities associated with the geodesic, which can be used  to simplify the equations of motion of the particle.
The motion of a particle with energy $E$ and angular momentum $l$ is governed by the equation
\begin{eqnarray}
\dot{r}^2+V(r)=\frac{1}{b^2}  \,,
\end{eqnarray}
where $b=l/E$ is the impact parameter, and
\begin{eqnarray}
V(r)=\left(\frac{l^2}{r^2}+\epsilon \right)f(r)
\end{eqnarray}
is the effective potential. Here, $\epsilon=0, 1$ for massless and massive particle respectively, and $``\cdot"$ denotes the derivative with respect to the parameter of the geodesic. Of course, this parameter is the so-called affine parameter when the particle is massless.

\begin{figure}[htb]
  \centering
\includegraphics[width=9cm]{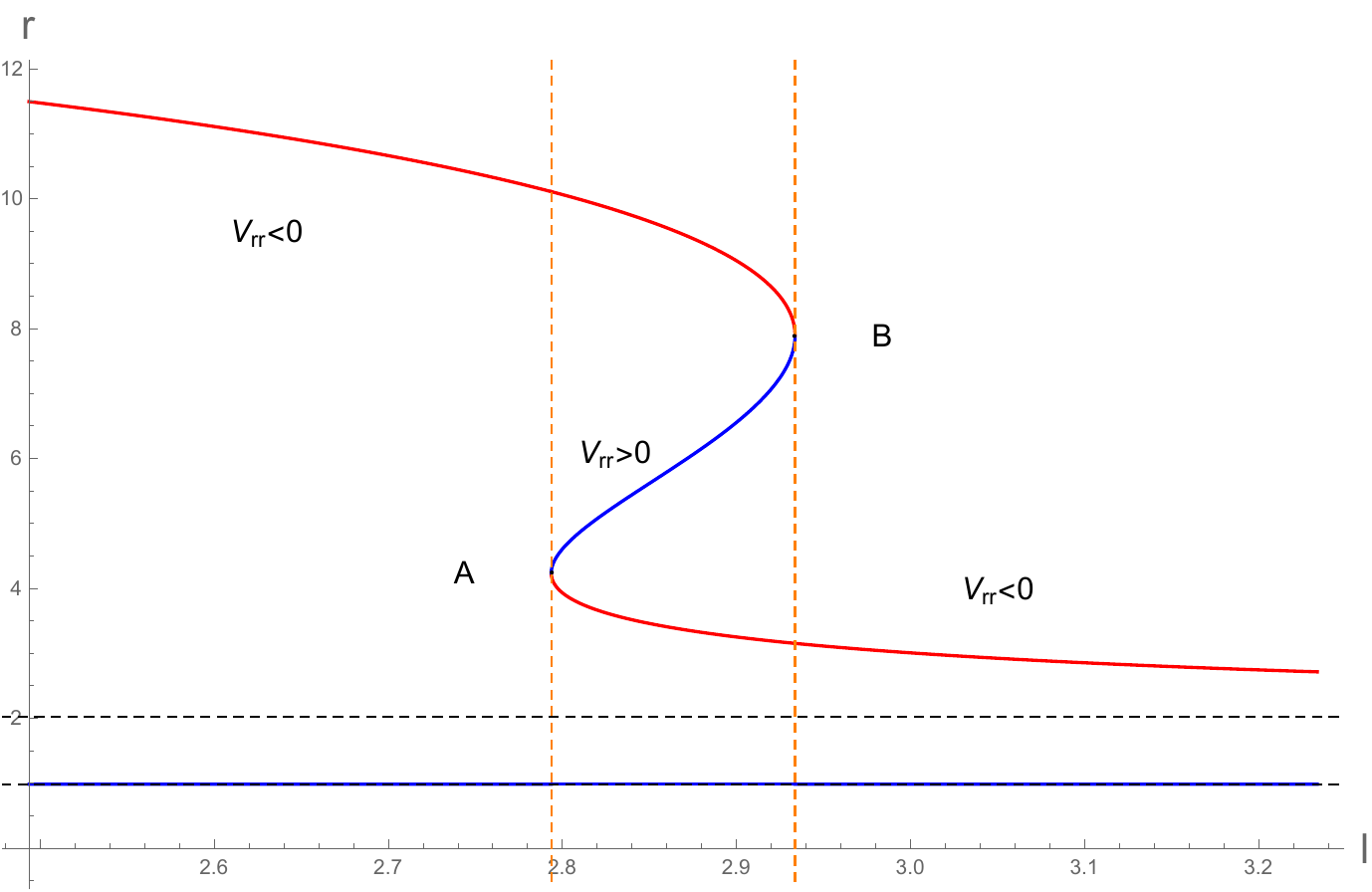}
  \caption{The $r-l$ diagram.
  The RN-dS black hole with $Q=0.995, \Lambda=0.001$.
  $l$ is the angular momentum of a massive particle and  $r$ is the corresponding radius of the circular orbital radius.
  Two horizontal black dashed lines are photon spheres,
  while the two solid curves in the figure are derived from $V_r=0$.
  The below one (blue), which is inside the event horizon, nearly overlaps with the black dashed line.
  The curve above is divided into three parts by points $A$ and $B$.
  The part $AB$ (blue) represents the stable circular orbit because $V_{rr}>0$ and the red curves represent the unstable circular orbit because $V_{rr}<0$.
  Therefore $A$ is the ISCO and $B$ is the OSCO.
  The two orange vertical dashed line, whose abscissas are $l_A$ and $l_B$, represent the angular momentum $l$ for point $A$ and $B$. }
  \label{rl}
\end{figure}

For the circular orbits, we have $V_r=0$, where $V_r$ denotes the derivative of $V$ with respect to $r$. Similarly, in the following discussion, $V_{rr}$ and $V_{rrr}$ represent the second and third derivative of $V$ with respect to $r$ respectively.
Furthermore, $V_{rr}<0$  for the unstable circular orbits and $V_{rr}>0$  for the stable circular orbits.
Solving $V_{r}=0$ for the massless particles, we can get a stable photon sphere between the event horizon and the Cauchy horizon, and an unstable photon sphere outside the event horizon.
However, in the case of massive particles, the situation is  complicated.
Fig.\ref{rl} represents the parameter space of circular orbits, where each point corresponds to a unique circular orbit.\footnote{Here and below, unless otherwise stated, all numerical values of $Q, r$, and $l$ represent quantities in units of $M$.
Similarly, $\Lambda$ is given in units of $M^{-2}$.}
The curves depicted in the figure is composed of points that satisfy $V_{r}=0$, representing physically allowed circular orbits.
The curves consist of two parts.
The lower part, appearing almost horizontal, is concealed behind the event horizon, thus is not of interest to us.
However, the stability of a point on the curve requires further calculations on $V_{rr}$.
As the radius $r$ approaches positive infinity or $0$, the derivative of effective potential $V_{r}$ tends towards negative infinity.
The $V_{r}$ within the region between the two branches of the curve is positive.
Intuitively, consider an auxiliary vertical line with fixed $l$.
The intersections of the line with the curve represent the zeros of $V_{r}$.
As we gradually increase the value of $r$ from bottom to top, crossing the intersections of the line and the curve, if $V_{r}$ transits from negative to positive, then $V_{rr}\geq0$.
Conversely, if $V_{r}$ transits from positive to negative, then $V_{rr}\leq0$.
A more rigorous discussion appears in the following paragraphs.

\begin{figure}[htb]
  \centering
\subfigure[]{{\includegraphics[width=5cm]{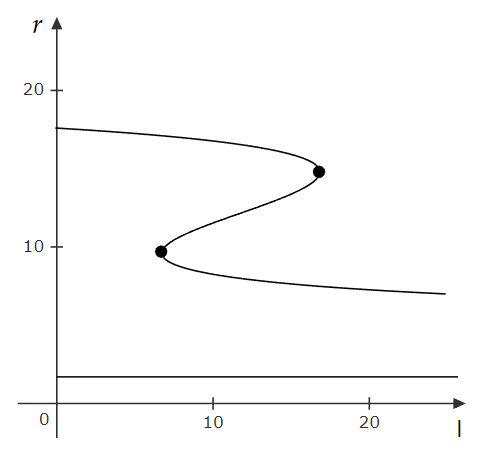}} \label{rl31}}
\subfigure[]{{\includegraphics[width=5cm]{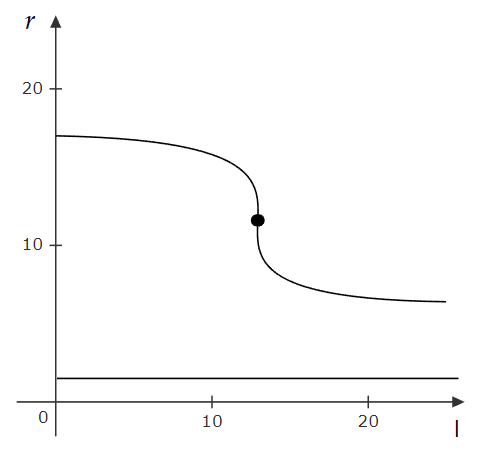}} \label{rl32}}
\subfigure[]{{\includegraphics[width=5cm]{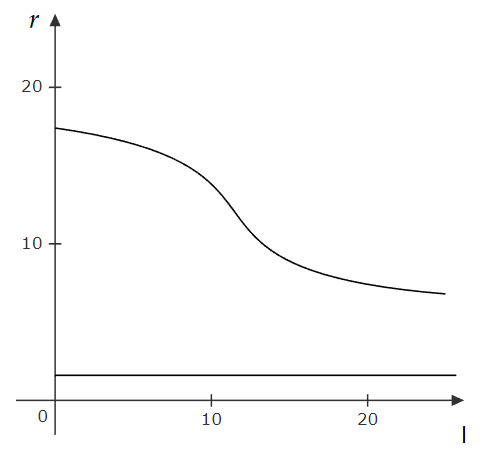}} \label{rl33}}
  \caption{Some possible $r-l$ diagram.
  The RN-dS black hole have an ISCO and an OSCO in figure (a).
  The ISCO and OSCO is coincide in figure (b).
  And there is no ISCO or OSCO in figure (c). }
\end{figure}

We have two special circular orbits represented by point $A$ and $B$ where $\mathrm{d} l/\mathrm{d} r=0$.  Actually, the circular orbit associated with $A$ where $\mathrm{d}^2 l/\mathrm{d} r^2>0$ is nothing but the ISCO. Accordingly, the orbit presented by the point $B$ where $\mathrm{d}^2 l/\mathrm{d} r^2<0$ is the OSCO mentioned in the previous section.
It should be noted that although neither point $A$ nor point $B$ represent stable circular orbits since $V_{r}=V_{rr}=0$, $V_{rrr}\neq0$ at these points, stable orbits can indeed be found within any small vicinity around them.
We will prove this in the next paragraph.
Besides the situation depicted in Fig.\ref{rl} or Fig.\ref{rl31}, there is a critical case where ISCO and OSCO are coincided, which has been shown in Fig.\ref{rl32}. It can be seen that $\mathrm{d}^2l /\mathrm{d} r^2$ is vanished on the black spot of Fig.\ref{rl32}. Unlike the usual Schwarzschild black hole, for some range of parameters, RN-dS spacetime does not possess stable circular orbits outside the event horizon, which can be understood from Fig.\ref{rl33}.
Obviously, as $l\rightarrow\infty$, the effect potential $V$ of the massive particles tends to that of the massless particles.
Meanwhile, the two curves of $r-l$ in Figs. \ref{rl31}, or \ref{rl32} and \ref{rl33}, tend to the photon spheres as $l\rightarrow\infty$.

The ISCO and OSCO can be obtained by solving the equations $V_r=0$ and $V_{rr}=0$. This is actually the usual definition of the stable circular orbits in literature \cite{Berry:2020ntz,Boonserm:2019nqq,Song:2021ziq}.
The $r-l$ curve in Fig.\ref{rl} is the solution of $V_r=0$. Denote the parameter of the upper branch of the curve as $\sigma$. Consider the $V$ as a function of $r$ and $l$. Since each $r$ uniquely corresponds to a point on the curve, we can take the $\sigma$ varies monotonically  with $r$, which further implies that $\mathrm{d} r/\mathrm{d} \sigma\neq 0$.
We have
\begin{eqnarray}
0=\frac{\mathrm{d} V_{r}\big(r(\sigma),l(\sigma)\big)}{\mathrm{d}\sigma}=\frac{\partial V_{r}}{\partial r}\frac{\mathrm{d} r}{\mathrm{d} \sigma}+\frac{\partial V_{r}}{\partial l}\frac{\mathrm{d} l}{\mathrm{d}\sigma}=V_{rr}\frac{\mathrm{d} r}{\mathrm{d} \sigma}+\frac{\partial V_{r}}{\partial l}\frac{\mathrm{d} l}{\mathrm{d}\sigma}
\end{eqnarray}
on the curve.
Thus
\begin{eqnarray}
V_{rr}=-\frac{\partial V_{r}}{\partial l}\frac{\mathrm{d} l}{\mathrm{d}r} =0  \,,
\end{eqnarray}
where we have used $\mathrm{d} l/\mathrm{d} r=0$ at points $A$ and $B$.
Therefore, the turning points $A$ and $B$ in Fig.\ref{rl} are satisfied by $V_r=V_{rr}=0$, which means they are actually the ISCO and OSCO.

The boundary of the range of parameters in which ISCO and OSCO exist is the critical case as Fig.\ref{rl32}, where ISCO and OSCO coincide.
In this case $V_{rrr}=0$ and this condition is equivalent to $\mathrm{d}^2 l/\mathrm{d} r^2=0$.
The proof is as follows.
At the point coincided by $A$ and $B$, $V_{rr}=0$ and $\mathrm{d}l/\mathrm{d}r=0$.
If we chose $\mathrm{d}r/\mathrm{d} \sigma$ is positive and finite, then
\begin{eqnarray}
\frac{\mathrm{d}l}{\mathrm{d}\sigma}=\frac{\mathrm{d}l}{\mathrm{d}r}\frac{\mathrm{d}r}{\mathrm{d}\sigma}=0 \,.
\end{eqnarray}
From
\begin{eqnarray}
0=\frac{\mathrm{d}^2V_{r}}{\mathrm{d}\sigma^2}
=V_{rrr} \left( \frac{\mathrm{d}r}{\mathrm{d} \sigma} \right)^2 +2\frac{\partial V_{rr}}{\partial l}\frac{\mathrm{d} l}{\mathrm{d} \sigma}\frac{\mathrm{d} r}{\mathrm{d} \sigma}
+\frac{\partial V_{r}}{\partial l}\frac{\mathrm{d}^2 l}{\mathrm{d} \sigma^2}+\frac{\partial^2 V_{r}}{\partial l^2}\left( \frac{\mathrm{d} l}{\mathrm{d} \sigma} \right)^2
+V_{rr}\frac{\mathrm{d}^2 r}{\mathrm{d} \sigma^2}
  \,,
\end{eqnarray}
we get
\begin{eqnarray}
V_{rrr}=-2\frac{\partial V_{rr}}{\partial l}\frac{\mathrm{d} l}{\mathrm{d} r}
-\frac{\partial V_{r}}{\partial l}\frac{\mathrm{d}^2 l}{\mathrm{d} \sigma^2}\left( \frac{\mathrm{d} r}{\mathrm{d} \sigma} \right)^{-2}
-\frac{\partial^2 V_{r}}{\partial l^2}\left( \frac{\mathrm{d} l}{\mathrm{d} r} \right)^2
-V_{rr}\frac{\mathrm{d}^2 r}{\mathrm{d} \sigma^2}\left( \frac{\mathrm{d} r}{\mathrm{d} \sigma} \right)^{-2}
=-\frac{\partial V_{r}}{\partial l}\frac{\mathrm{d}^2 l}{\mathrm{d} \sigma^2}\left( \frac{\mathrm{d} r}{\mathrm{d} \sigma} \right)^{-2}\,,
\end{eqnarray}
where we have used $\mathrm{d} l/\mathrm{d} r=0$ and $V_{rr}=0$ for ISCO and OSCO, and $V_{r}=0$ for the entire curve in Fig.\ref{rl32}.
Hence when $V_{rrr}=0$ and $\partial V_r/\partial l\neq 0$, we have $\mathrm{d}^2 l/\mathrm{d} \sigma^2=0$.
And then
\begin{eqnarray}
\frac{\mathrm{d}^2 l}{\mathrm{d} r^2}
=\frac{\mathrm{d}^2 l}{\mathrm{d} \sigma^2}\left( \frac{\mathrm{d} r}{\mathrm{d} \sigma} \right)^{-2}
-\frac{\mathrm{d} l}{\mathrm{d} \sigma}\left( \frac{\mathrm{d} r}{\mathrm{d} \sigma} \right)^{-3}\frac{\mathrm{d}^2 r}{\mathrm{d} \sigma^2}=0  \,.
\end{eqnarray}
Therefore, when $V_{rrr}=V_{rr}=0$ and $\mathrm{d}l/\mathrm{d}r=0$, we have $\mathrm{d}^2 l/\mathrm{d} r^2=0$.
To summarize, the orbit radii of ISCO and OSCO are satisfied by $V_{r}=V_{rr}=0$.
And the critical case where ISCO and OSCO coincide, which is also the critical case that there is no ISCO and OSCO, is satisfied by $V_{r}=V_{rr}=V_{rrr}=0$.

\begin{figure}[htb]
  \centering
\subfigure[]{{\includegraphics[width=5cm]{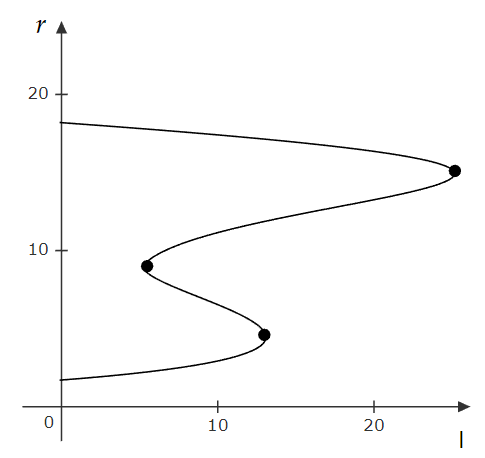}} \label{rl41}}
\subfigure[]{{\includegraphics[width=5cm]{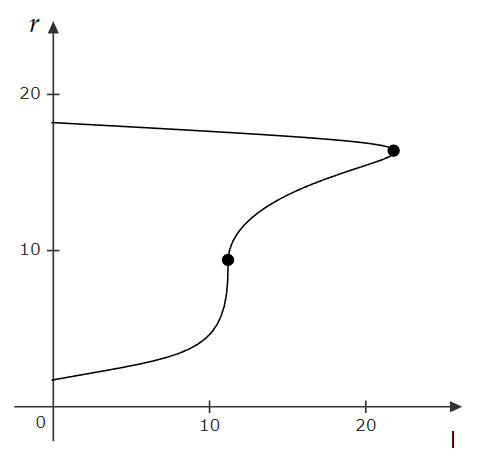}} \label{rl42}}
\subfigure[]{{\includegraphics[width=5cm]{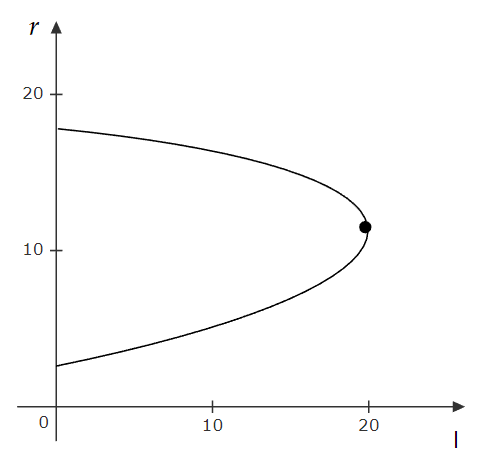}} \label{rl43}}
  \caption{Some possible $r-l$ diagrams without photon sphere.
  The spacetime have two ISCOs and two OSCOs in figure (a) (the three turning points and an ISCO with $l=0$).
  An ISCO and OSCO is coincide in figure (b). And there is an ISCO ($l=0$) and an OSCO (the turning point) in this case.
  There is an ISCO ($l=0$) and an OSCO (the turning point) in figure (c). }
  \label{rl4}
\end{figure}

If there is no photon sphere, i.e., the spacetime is the RN-dS naked singularity, the $r-l$ diagrams will be different, and are shown in Fig.\ref{rl4}.

\begin{figure}[htb]
  \centering
  \includegraphics[width=9cm]{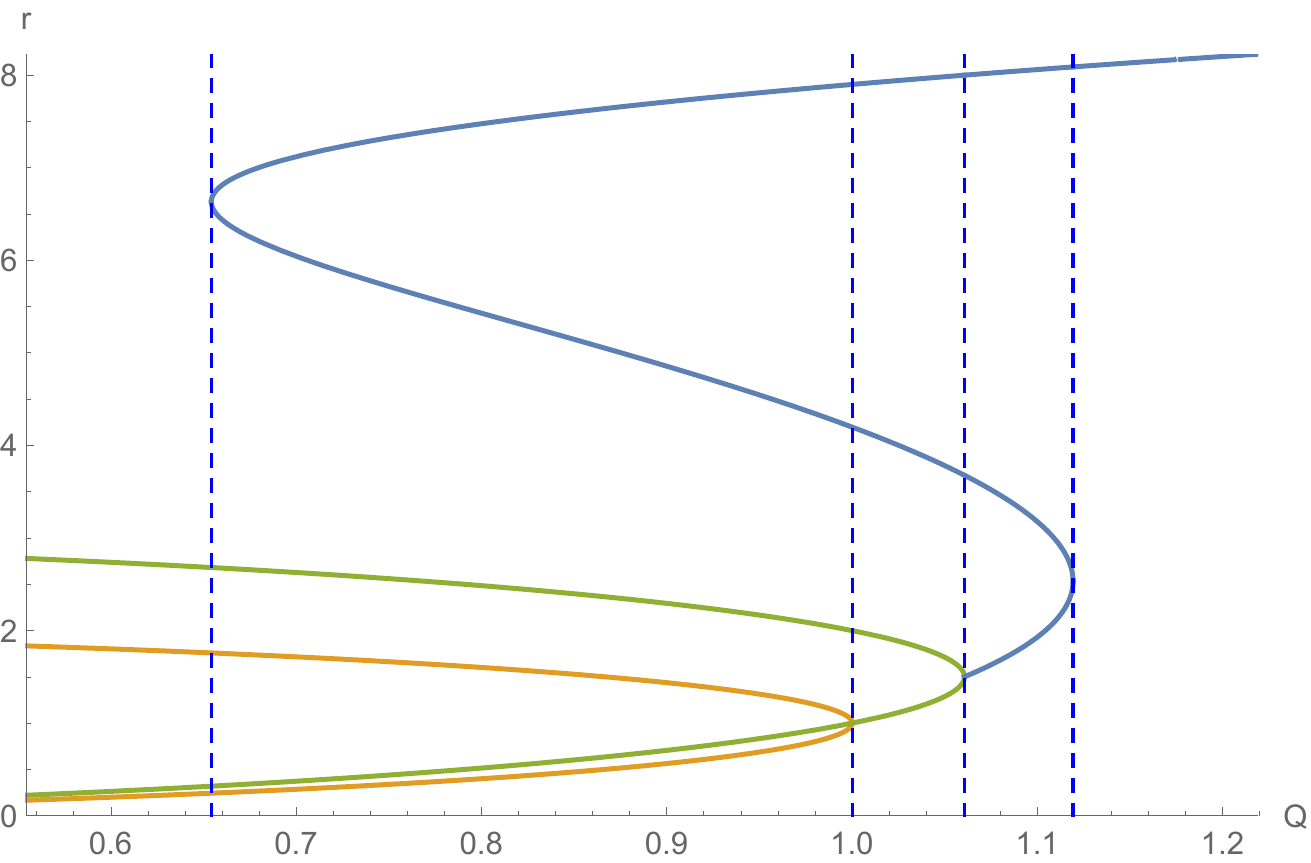}
  \caption{The blue curve is solution of $V_{r}=V_{rr}=0$, i.e., the ISCO or OSCO.
  The orange curve is the Cauchy horizon and the event horizon (The radius of the cosmological horizon is too large to be shown in the figure, so we  omit it).
  The green curve is the photon spheres.
  The parameter of the RN-dS spacetime is $\Lambda=0.001$.
  The four blue dashed line passing the turning points are located at $Q_1=0.654242, Q_2=1.00017, Q_3=1.06066, Q_4=1.11927$.}
  \label{rQ}
\end{figure}

\begin{figure}[htb]
  \centering
\subfigure[]{{\includegraphics[width=6cm]{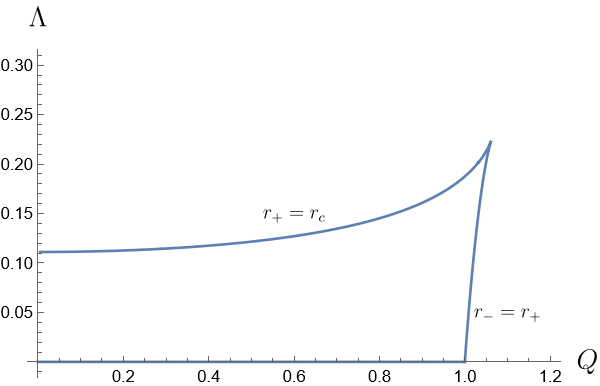}} \label{range1}}
\subfigure[]{{\includegraphics[width=6cm]{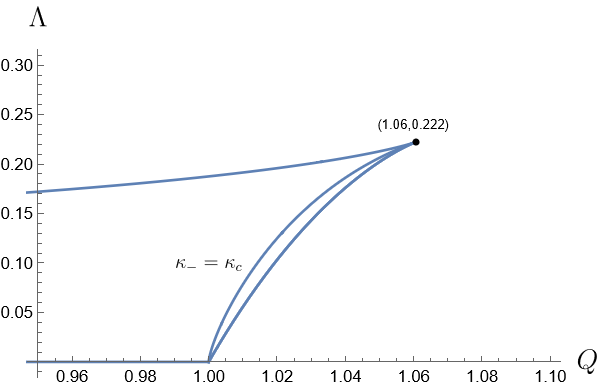}} \label{range2}}
  \caption{Figure (a) is the range of the parameter of the RN-dS black hole with three horizons.
  The boundary of the range is $r_-=r_+$ and $r_+=r_\mathrm{c}$.
  Figure (b) is the range of the parameter of the RN-dS black hole with $\kappa_-<\kappa_\mathrm{c}$.
  The boundary is $\kappa_-=\kappa_\mathrm{c}$ and $\kappa_-=0$.}
\end{figure}

In order to see the various scenarios clearly, the relation of the existence of ISCO and OSCO and $Q$ is shown in Fig.\ref{rQ}.
As we can see, the unstable photon sphere is always outside the event horizon. The dependence  on the charge $Q$ is given as follows
\begin{itemize}
\item[(i).]For $Q\in(0,Q_2)$, there are three horizons and two photon spheres.
There is no ISCO and OSCO for $Q\in(0,Q_1)$, as Fig.\ref{rl33}, and an ISCO and an OSCO for $Q\in(Q_1,Q_2)$, as Fig.\ref{rl31}.
When $Q=Q_1$, the ISCO and OSCO coincide, as Fig.\ref{rl32}.
\item[(ii).] For $Q\in(Q_2,Q_3)$, it is a RN-dS naked singularity with two photon spheres, and there is an ISCO and an OSCO as Fig.\ref{rl31}, too.
\item[(iii).]For $Q\in(Q_3,Q_4)$, the RN-dS naked singularity has no photon sphere and the blue curve has three turning points.
 Stable circular orbits exist for two distinct intervals of $r$ in this case, as Fig.\ref{rl41}.
\item[(iv).] When $Q\geqslant Q_4$, there is an ISCO and an OSCO, and the corresponding $r-l$ diagrams are Fig.\ref{rl42} for $Q=Q_4$ and Fig.\ref{rl43} for $Q>Q_4$.
\end{itemize}
The stable Cauchy horizon emerges when $\kappa_-<\kappa_\mathrm{c}$ \cite{Brady:1992,Cai:1995ux}, where $\kappa_-$ and $\kappa_\mathrm{c}$ are the surface gravities of the Cauchy horizon and cosmological horizon respectively. Once a stable Cauchy horizon exists,
the SCCC will be broken down and the predictability of classical theory is threatened.
Thus it aroused great interest in the community of general relativity and gravity theory.
We will look for the range of the parameters of the black hole which has a stable Cauchy horizon as well as stable circular orbits (i.e., ISCO and OSCO).
Firstly, the range of the parameters for the black hole with three horizons has been drawn in Fig.\ref{range1}. A part of the boundary of this range is given by $r_+=r_-$ and $r_+=r_{\mathrm{c}}$.
The parameter range for Kerr-dS black holes can be found in \cite{Momennia:2023lau}.
Although their graph closely resembles ours, it is important to note that their abscissa represents the angular momentum parameter $\alpha$, while our abscissa corresponds to the charge parameter $Q$.
This observation highlights the similarity between angular momentum and charge, illustrating that the RN-dS black hole can serve as a valuable reference for further study on Kerr-dS black holes.
Secondly, the range of the parameters of the RN-dS black hole with a stable Cauchy horizon is shown as the small crescent in Fig.\ref{range2}.
The left boundary of this area is $\kappa_-=\kappa_\mathrm{c}$ and the right boundary is $\kappa_-=0$, i.e., $r_-=r_+$.

\begin{figure}[htb]
  \centering
  \includegraphics[width=16cm]{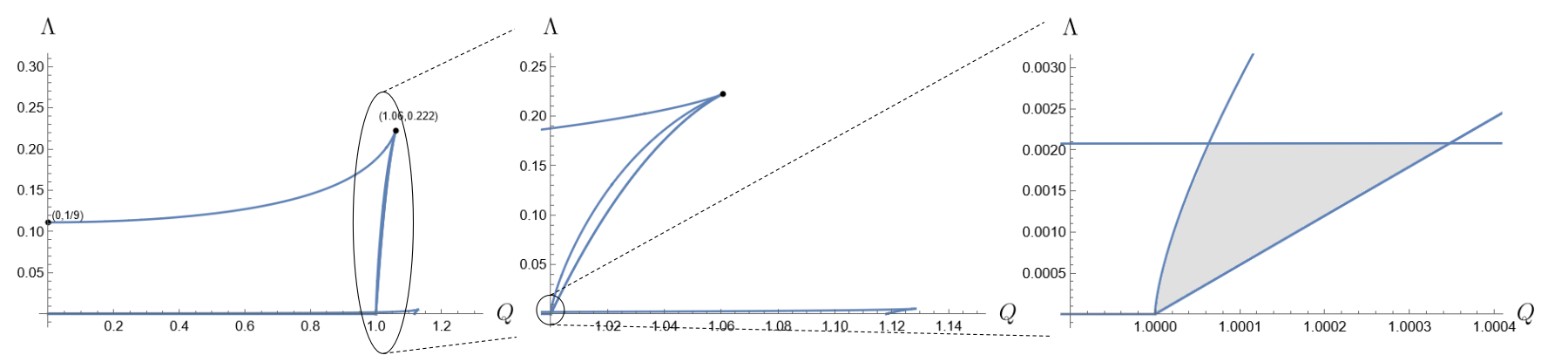}
  \caption{The left figure is the range where there are three horizons, one of which is a stable Cauchy horizon, and an ISCO and OSCO.
  The middle figure and the right figure are the partial enlarged figure the left one.
  The range is marked as a gray area in the right figure. }
  \label{range4}
\end{figure}

Finally, the range of parameters for the presence of  stable circular orbits, i.e., ISCO and OSCO, has been found from the definition the result is put in Fig.\ref{range4}.
In summary, the final range is the gray area in the right figure of Fig.\ref{range4}.

\section{The appearance of the Schwarzschild-dS black hole viewed by observers with different distance  }\label{S3}
\begin{figure}[htb]
  \centering
  \includegraphics[width=10cm]{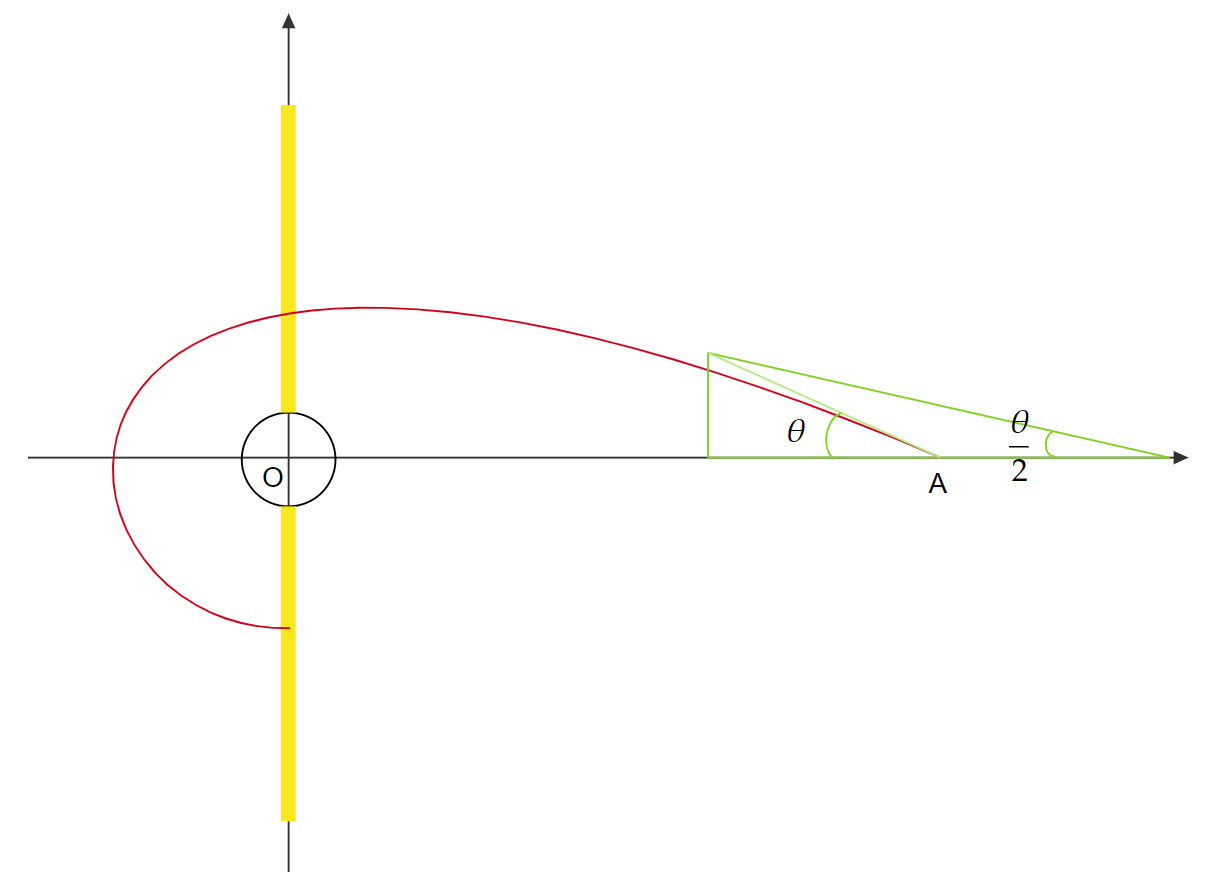}
  \caption{The schematic picture showing the trajectory of the photons and the stereographic projection.
  The black hole is located at $O$ and the observer is at $A$.
  The black ring is the event horizon of the black hole.
  The yellow line represents the accretion disk and the red curve is the ray from the accretion disk to the observer.
  $\theta$ is the angle of incidence.
  The green triangle is a schematic diagram of the stereographic projection.}
  \label{traj}
\end{figure}

In this section we will study the effect of the position of the observer on the image of the Schwarzschild-dS black hole.
The positions of the black hole, accretion disk, observer and the trajectory of the photons are shown in Fig.\ref{traj}.
We use the stereographic projection to get the image, and thus the abscissa of the image $y_\mathrm{p}$ \cite{Grenzebach_fort} is
\begin{eqnarray}
y_\mathrm{p}=2\tan\frac{\theta}{2}=2\left.\left( \frac{1}{\sqrt{f}}\frac{r}{b}-\sqrt{\frac{r^2}{fb^2}-1} \right) \right|_{r=r_\mathrm{obs}}  \,,
\end{eqnarray}
where $b$ is the impact parameter and $r_\mathrm{obs}$ can be understood as the distance between the observer and the black hole.

We use the ray-tracing method \cite{Gralla:2019xty} to draw the image of the black holes.
The normalized number of orbits $n=\phi/(2\pi)$ relates to the number of intersections with the equatorial plane of a particular light ray, where $\phi$ is the azimuthal angle.
Light bends greatly around the massive object, some even trace a circular path at the peak of the effective potential. This circular path also known as the photon sphere or ``critical curve".
At the critical impact parameter, the number of rotations of the photons increases infinitely.
This results in a bright ring in the image, and the size of the ring only depends on the background geometry of spacetime.
Images produced by rays intersecting the accretion disk once, twice, or more than twice are referred to as ``direct emission", ``lensed ring", and ``photon ring", respectively.
However, in most cases, the photon ring is so close to the lensed ring that it can not be distinguished, and its contribution to the overall intensity is negligible.
It means we can see only one bright ring in the image.
At each intersection, newly emitted photons from the accretion disk join in the journey towards the screen or the celestial sphere of the observer.
Each of intersections of light with the accretion disk contributes to the intensity received by the observer.
Besides, considering the effect of gravitational redshift on the intensity of the emission, the intensity of the light received by the observer, $I_\mathrm{obs}$, is \cite{Gralla:2019xty,DeMartino:2023ovj,Rybicki:2004hfl}
\begin{eqnarray}
\label{Iobs}
I_\mathrm{obs}=\sum_n I_\mathrm{em}(r)\left.\frac{f^2(r)}{f^2(r_\mathrm{obs})}\right|_{r=r_n}  \,,
\end{eqnarray}
where $r_n$ is the position of $n$-th intersection with the accretion disk.

\begin{figure}[htbp]
\centering
\subfigure[]{\includegraphics[width=4cm]{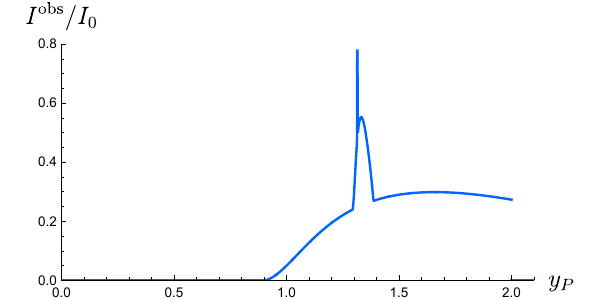}}
\subfigure[]{\includegraphics[width=4cm]{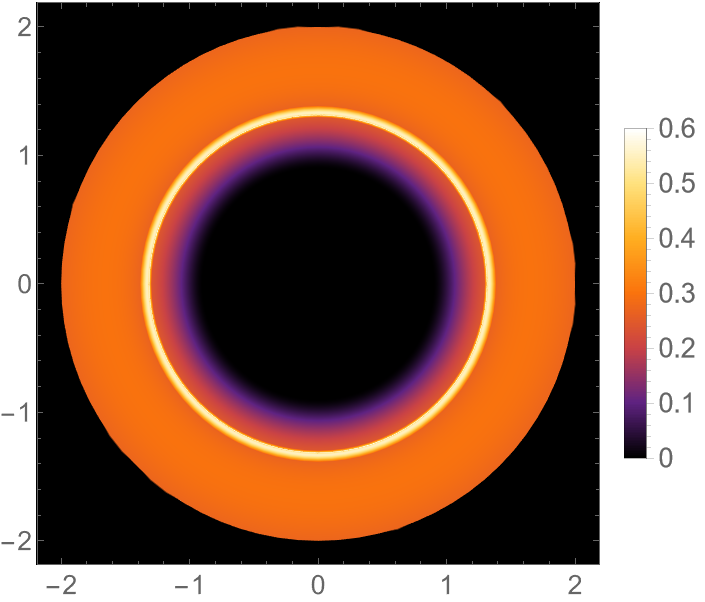}}
\subfigure[]{\includegraphics[width=4cm]{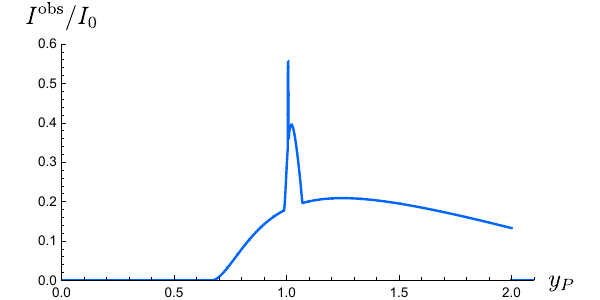}}
\subfigure[]{\includegraphics[width=4cm]{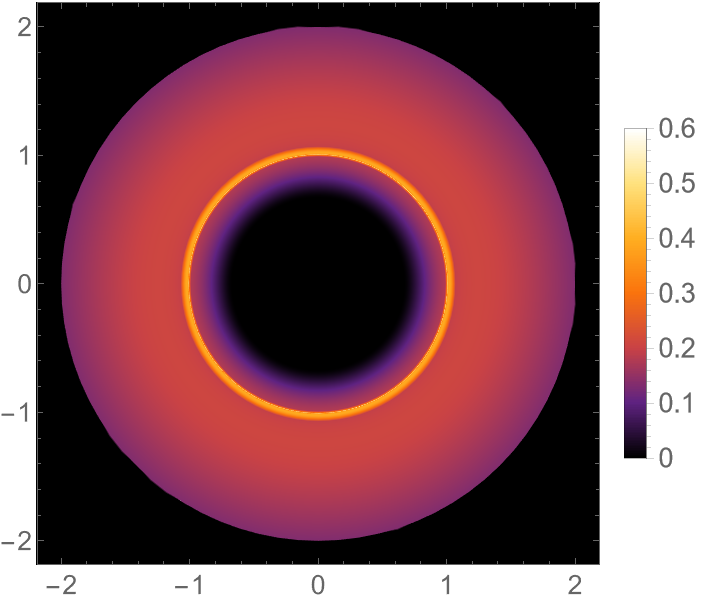}}

\subfigure[]{\includegraphics[width=4cm]{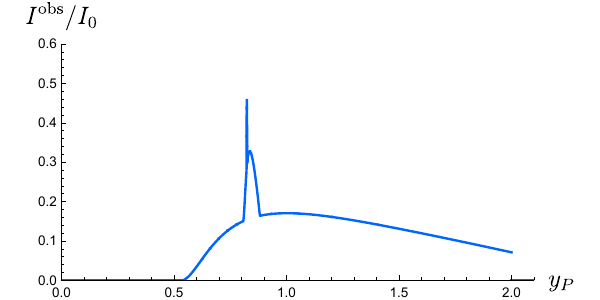}}
\subfigure[]{\includegraphics[width=4cm]{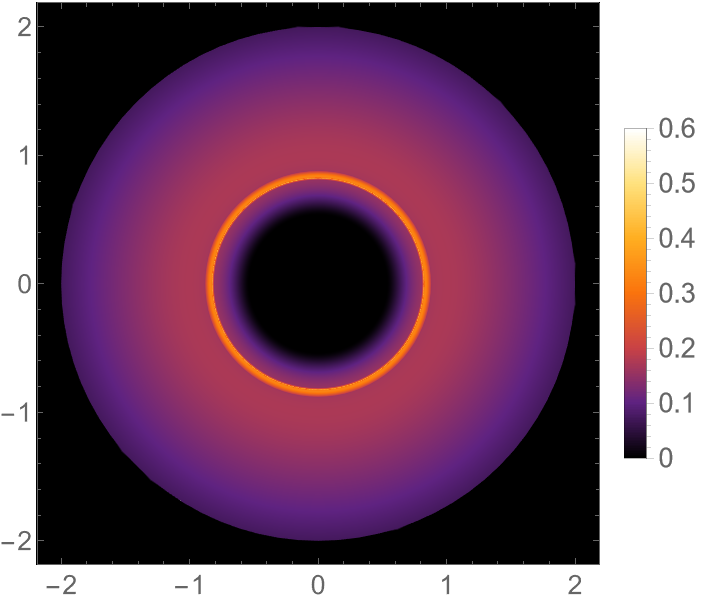}}
\subfigure[]{\includegraphics[width=4cm]{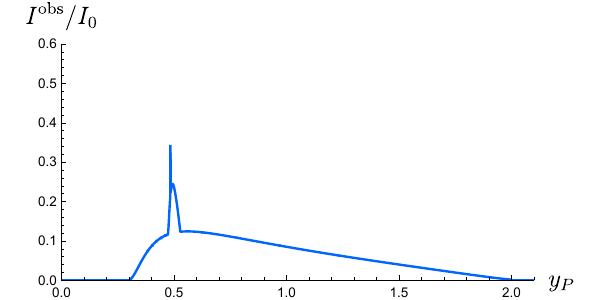}}
\subfigure[]{\includegraphics[width=4cm]{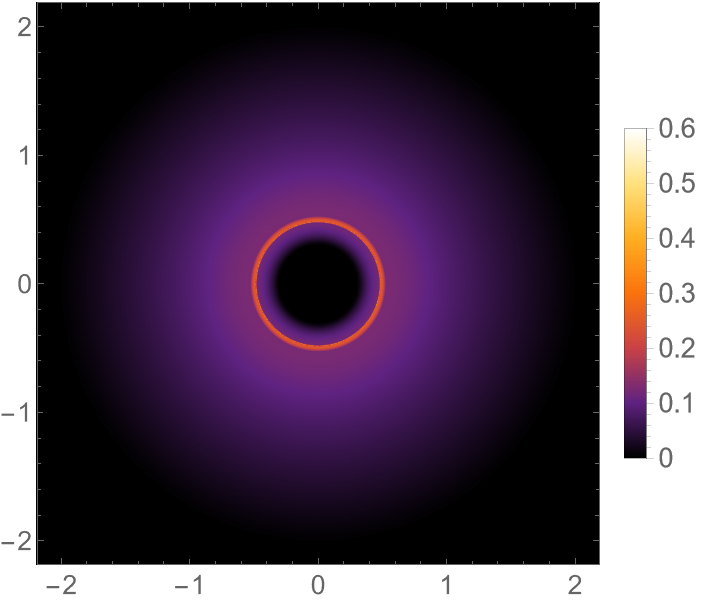}}

\subfigure[]{\includegraphics[width=4cm]{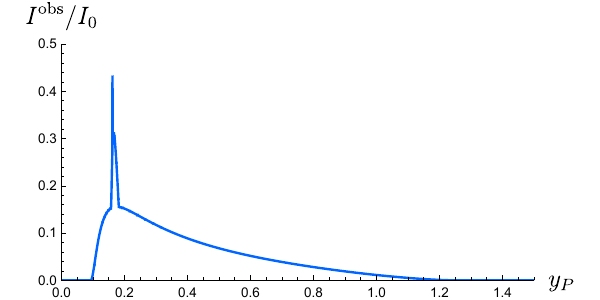}}
\subfigure[]{\includegraphics[width=4cm]{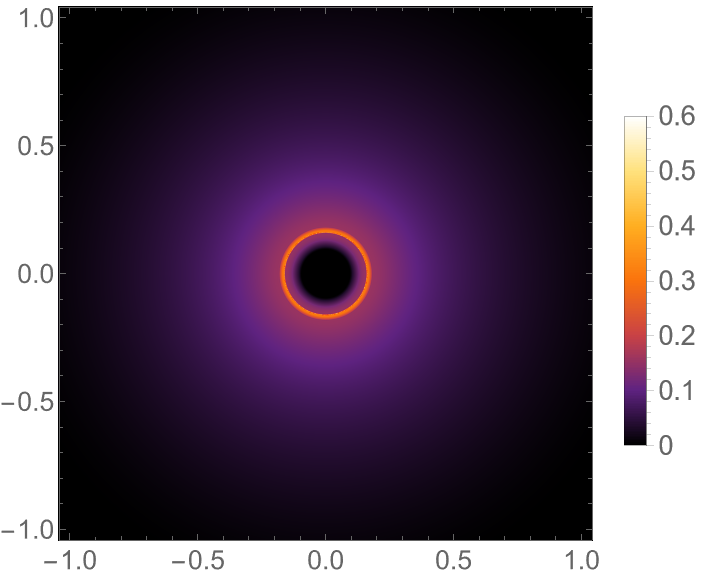}}
\subfigure[]{\includegraphics[width=4cm]{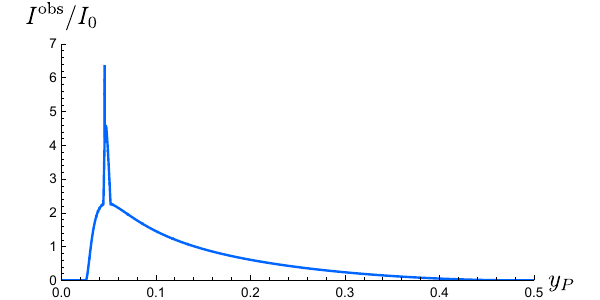}}
\subfigure[]{\includegraphics[width=4cm]{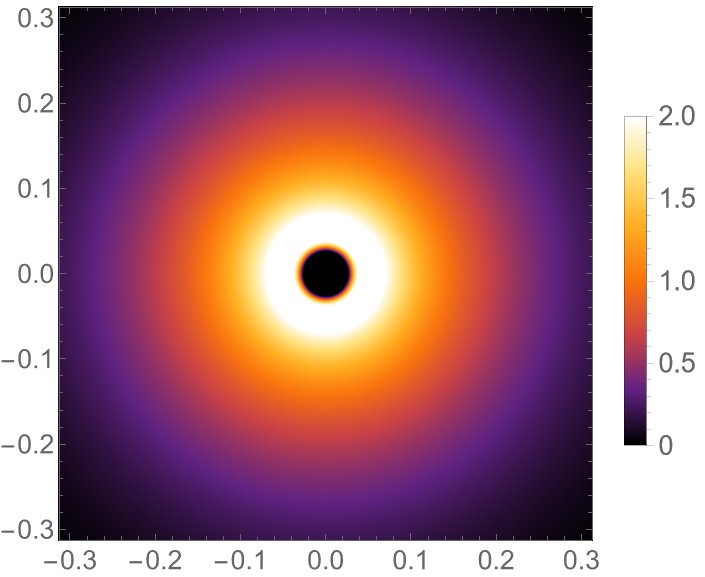}}
\caption{The appearance of the Schwarzschild-dS black hole with $\Lambda=0.001$ viewed by observers with different distance.
The observed intensity $I_\mathrm{obs}$ is normalized to the maximum value $I_0$ of the emitted intensity.
We choose six different observation positions, i.e., $r_\mathrm{obs}=4, 5, 6, 10, 0.5r_\mathrm{c}$ and $0.9r_\mathrm{c}$.
For each given observation position, we draw the corresponding observed intensity and shadow image separately.
Each item corresponds to an observer located at a specific position in a sequential manner.
(e.g., Figure (a) and (b) is the intensity and the image corresponds to observer at $r_\mathrm{obs}=4$, Figure (c) and (d) corresponds to observer at $r_\mathrm{obs}=5$, Figure (e) and (f) corresponds to observer at $r_\mathrm{obs}=6$, etc.)}
\label{SdS}
\end{figure}

A phenomenon of discontinuity arises when the location of the observer is finite, and the intensity received by the observer ``jumps" to null at certain impact parameter.
In order to see the ``jump'' of the $I_\mathrm{obs}$ caused by the position of the observer, we choose a Schwarzschild-dS black hole with $\Lambda=0.001 M^{-2}$, which has no ISCO and OSCO.
The luminous intensity of the accretion disk is
\begin{eqnarray}
\label{Iemrp}
I_\mathrm{em}(r) = \begin{cases}
\dfrac{\pi/2-\arctan (r-2)}{\pi/2-\arctan (r_+-2)} , & \text{if } r_+<r<r_c  \,.\\
~~0, & \text{if } r\leqslant r_+  \,\,\text{or}\, \, r\geqslant r_\mathrm{c}  \,.
\end{cases}
\end{eqnarray}
This kind of jump is different from the jump caused by the OSCO, which will be discussed later.
The image viewed by observers at different distance is given by Fig.\ref{SdS}.

\begin{figure}[htb]
  \centering
  \includegraphics[width=6cm]{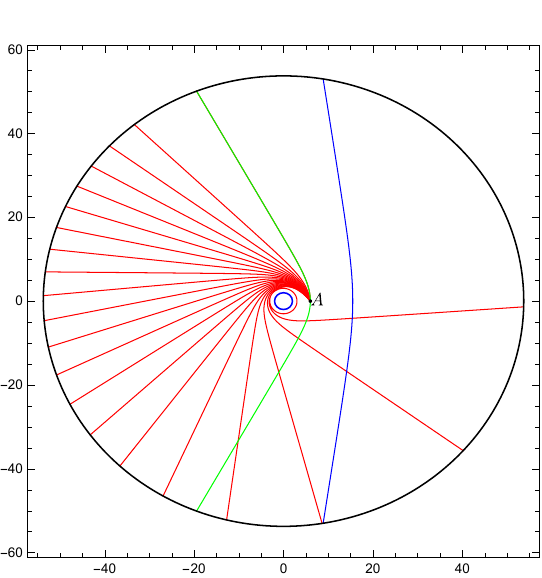}
  \caption{The trajectory of photons.
  The parameter of black hole is $\Lambda=0.001$.
  The blue ring is  the event horizon  and the black ring is the cosmological horizon.
  The observer is located at $A$.
  The rays with small $b$ are received by the observer, which are shown as red curves.
  But some rays with large $b$, whose radius of the perihelion is larger than $r_\mathrm{obs}$, can not pass through the observer.
  This kind of the rays is shown as the blue curve.
  And the critical case, when the perihelion of the ray is the position of the observer, is shown as the green} curve.
  \label{traj2}
\end{figure}

As we can see, there is a significant discontinuous jump in $I_\mathrm{obs}$ and an edge in the image when the observer is near the black hole.
This is because some rays with large $b$ can not be received by the observer located  too close to the black hole.
To explain this better, we draw the trajectories of photons in Fig.\ref{traj2}.
The rays with large $b$, like the blue curve, can not be received by the observer because its perihelion is farther than the position of the observer.
Therefore, there is a jump in the observed intensity.
And the image of the black hole has a significant edge if the observer is close enough to the black hole.
In order to avoid this kind of jump, the observer has to be located far way from the black hole.
In this case, only light rays with a significantly large value of $b$ cannot be received by the observer. Such light can only be emitted from the remote regions of the accretion disk. Consequently, its contribution to the total intensity is negligible.

\begin{figure}[htb]
  \centering
  \includegraphics[width=6cm]{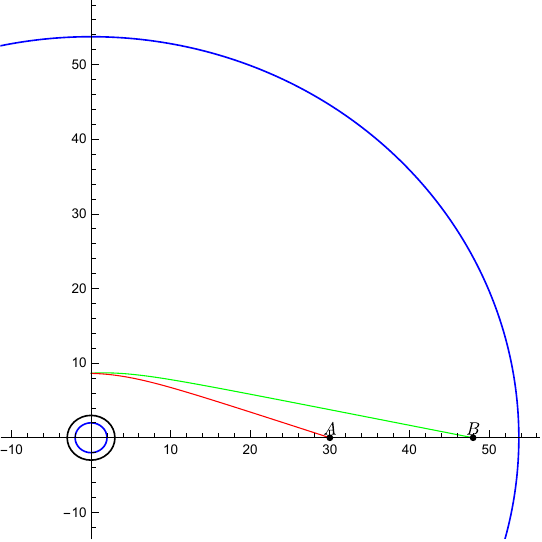}
  \caption{Two trajectories of rays from the accretion disk to the observers $A$ at $r_A=30$ and $B$ at $r_B=48$.
  The three rings are the Cauchy horizon, event horizon and cosmological horizon of the Schwarzschild-dS black hole with $\Lambda=0.001$ in turn.}
  \label{SdStraj}
\end{figure}

We also find that observer at $0.9r_\mathrm{c}$ receive a much stronger intensity $I_\mathrm{obs}$ compared to those  at other locations in Fig.\ref{SdS} since they have different redshift factor.
As shown in schematic diagram \ref{SdStraj}, although the two rays are emitted at the same position and they have same $f(r_\mathrm{em})$ and $I_\mathrm{em}(r_\mathrm{em})$, they have different $f(r_\mathrm{obs})$.
The two observers are located at $A$ and $B$ with $f_\mathrm{Aobs}=0.633$ and $f_\mathrm{Bobs}=0.190$, while $f_\mathrm{em}(r_\mathrm{em})=0.743$.
Therefore, from (\ref{Iobs}), the observed intensity $I_\mathrm{obs}$ for $A$ and $B$ varies by a factor of $f^2_\mathrm{Aobs}/f^2_\mathrm{Bobs}\thickapprox11$.
In conclusion, for observers located near the cosmological horizon, the intensity they received will become brighter as they get closer to the cosmological horizon. In fact, this intensity can reach infinity.
As the observer moves away from the black hole, the metric function $f(r_\mathrm{obs})$ increases and then decreases.
Due to the same reason, the intensity (\ref{Iobs}) is large at small $r_\mathrm{obs}$ and near the cosmological horizon, while diminishes a lot in the middle of these two horizons.

\begin{figure}[htbp]
\centering
\subfigure[]{\includegraphics[width=4cm]{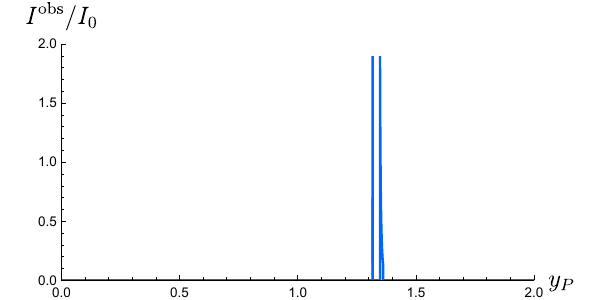}}
\subfigure[]{\includegraphics[width=4cm]{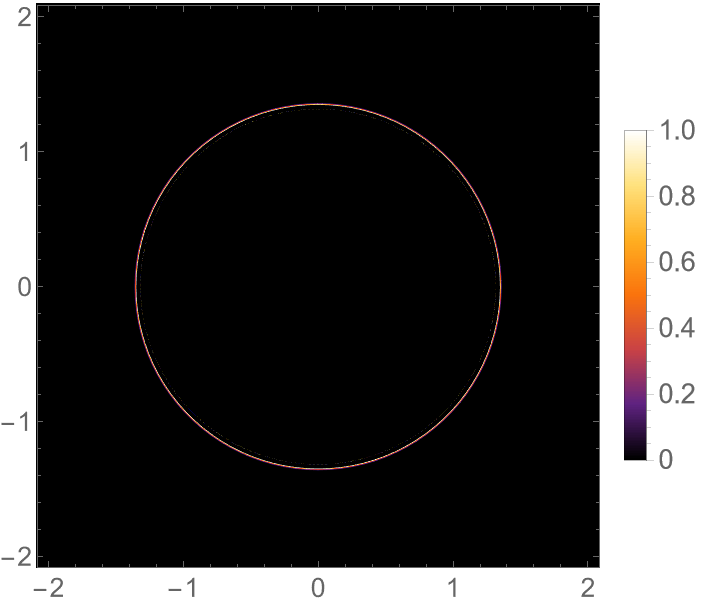}}
\subfigure[]{\includegraphics[width=4cm]{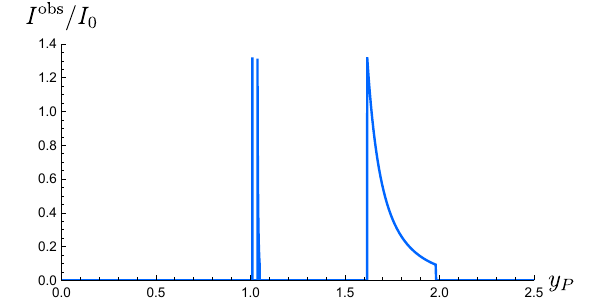}}
\subfigure[]{\includegraphics[width=4cm]{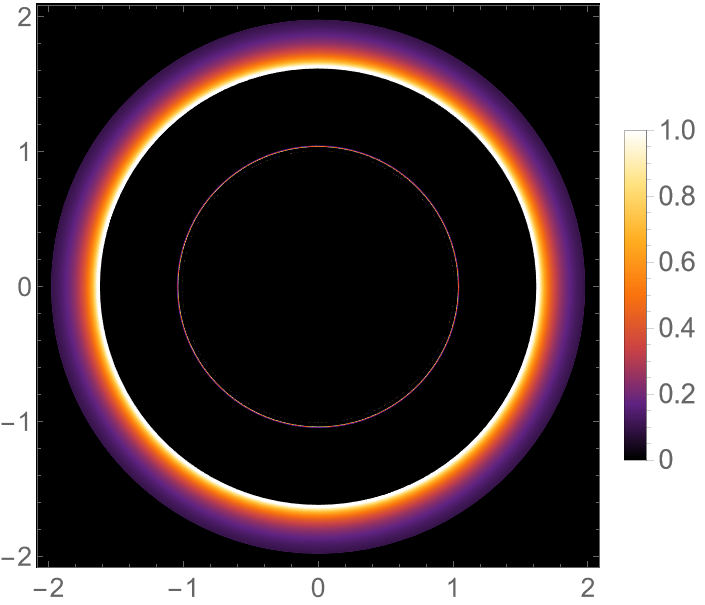}}

\subfigure[]{\includegraphics[width=4cm]{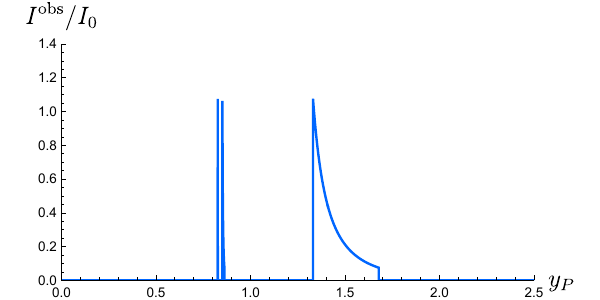}}
\subfigure[]{\includegraphics[width=4cm]{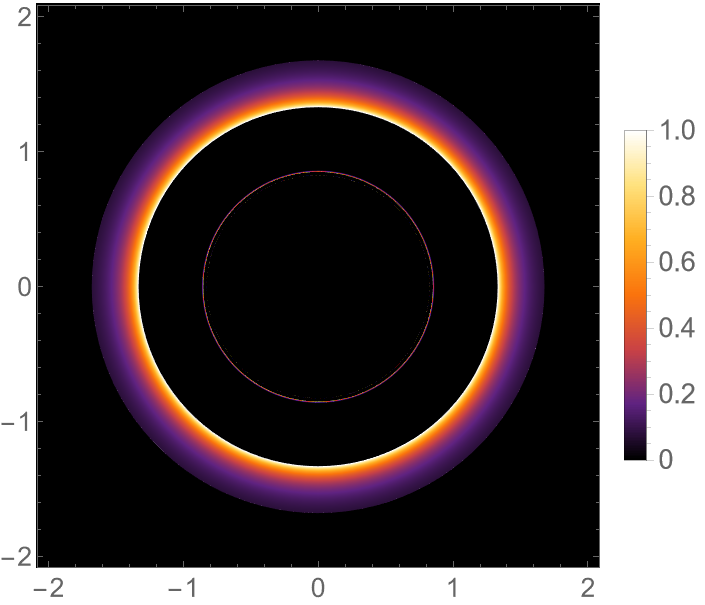}}
\subfigure[]{\includegraphics[width=4cm]{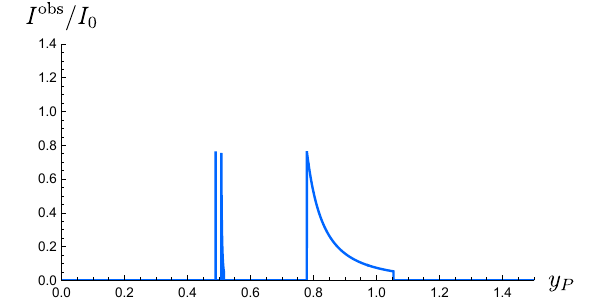}}
\subfigure[]{\includegraphics[width=4cm]{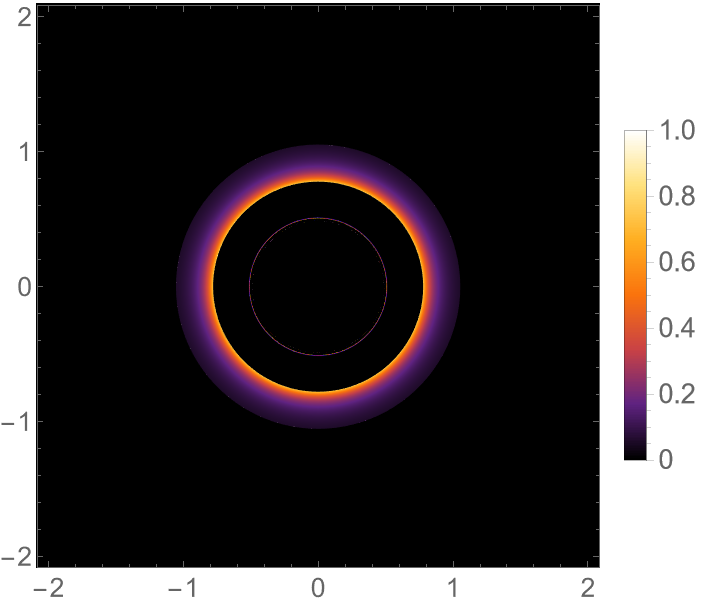}}

\subfigure[]{\includegraphics[width=4cm]{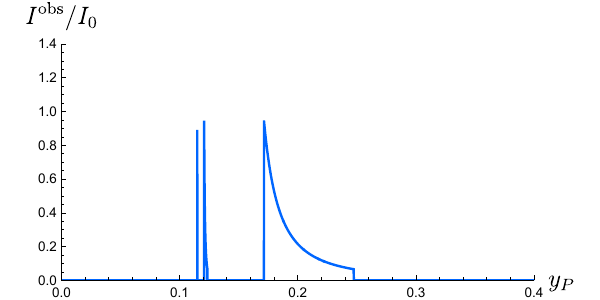}}
\subfigure[]{\includegraphics[width=4cm]{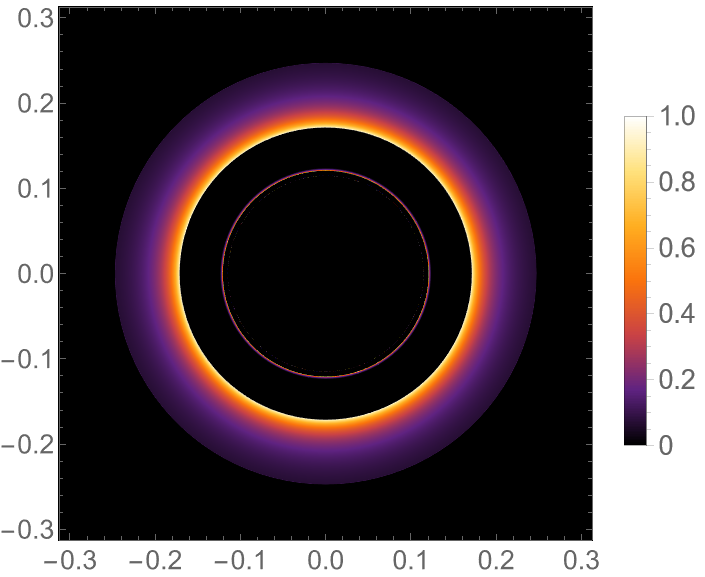}}
\subfigure[]{\includegraphics[width=4cm]{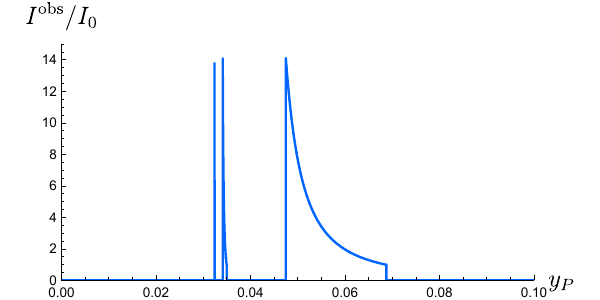}}
\subfigure[]{\includegraphics[width=4cm]{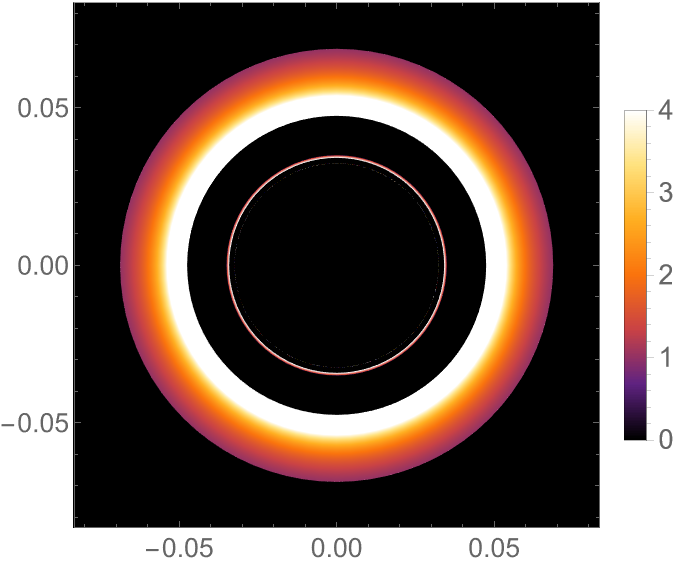}}

\caption{The appearance of the Schwarzschild-dS black hole with $\Lambda=0.0005$ viewed by observers with different distance.
It has an ISCO and an OSCO.
As the Schwarzschild-dS black hole without an ISCO and OSCO, we choose six different observation positions, i.e., $r_\mathrm{obs}=4, 5, 6, 10, 0.5r_\mathrm{c}$ and $0.9r_\mathrm{c}$.
(e.g., Figure (a) and (b) corresponds to observer at $r_\mathrm{obs}=4$, Figure (c) and (d) corresponds to observer at $r_\mathrm{obs}=5$, Figure (e) and (f) corresponds to observer at $r_\mathrm{obs}=6$, etc.}
\label{imageSdSIO}
\end{figure}

The image viewed by the observer far away from the Schwarzschild-dS black hole without ISCO or OSCO has the similar shape to the usual Schwarzschild black hole.
So we draw up the image of the Schwarzschild-dS black hole with an ISCO and OSCO with $\Lambda=0.0005$ in Fig.\ref{imageSdSIO}.
In this case, the accretion disk is distributed between ISCO and OSCO.
And the luminous intensity of the accretion disk is
\begin{eqnarray}
I_\mathrm{em}(r) = \begin{cases}
\dfrac{1}{(r-r_\mathrm{ISCO}+1)^2} , & \text{if }  r_\mathrm{ISCO}\leqslant r\leqslant r_\mathrm{OSCO} \\
0, & \text{if } r<r_\mathrm{ISCO} \quad \text{or } \quad r>r_\mathrm{OSCO}
\end{cases}    \,.
\label{IemrISCO}
\end{eqnarray}
To avoid the undesirable discontinuity due to the position of the observer, we set the observers as far as possible. Actually, in following discussion, $r_{\mathrm{obs}}=0.9r_{\mathrm{c}}$ can reach the requirement.

As in Fig.\ref{SdS},  the image of Schwarzschild-dS black hole with an ISCO and OSCO in Fig.\ref{imageSdSIO} also has a edge.
Particularly, when the observer is too close to the black hole, e.g., $r_\mathrm{obs}=4$, only the rays emitted from a little range of the accretion disk can be received by the observer.
In this case, despite being direct emission, the image contract to a ring.
However, the edges of the Schwarzschild-dS black hole with and without an ISCO and OSCO have different causes.
Besides the influence of the observed distant, the edge in the image of black hole with ISCO and OSCO is also resulting for that there is no light source outside the OSCO, i.e., the light source has a cutoff at OSCO.
Therefore, the image of the black hole with cosmological constant will be a little bit different from that of the  usual asymptotically flat Schwarzschild black hole.
The presence of the cosmological constant may lead to the emergence of OSCO, and this produce an edge in the image of the black hole.

\section{The appearance of the RN-dS black hole}\label{S4}
\begin{figure}[htb]
  \centering
  \includegraphics[width=6cm]{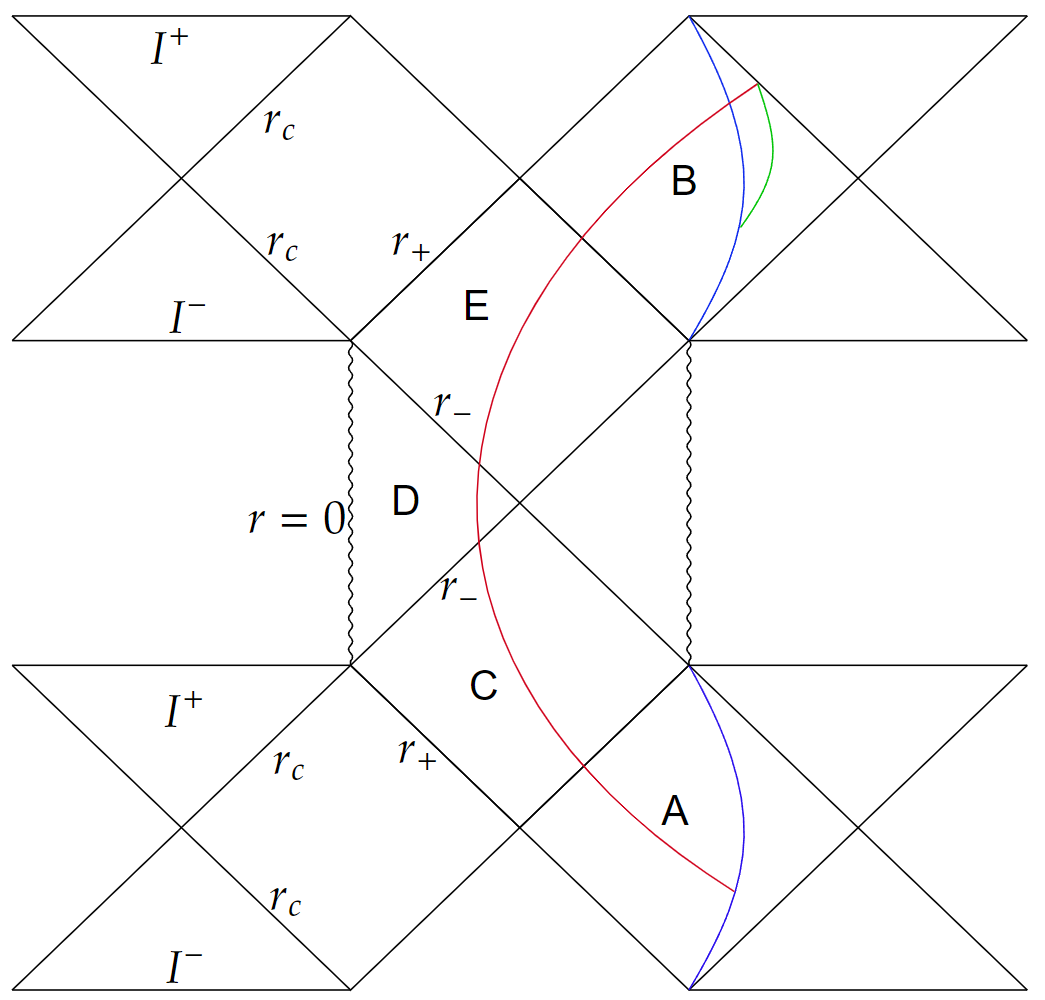}
  \caption{The Penrose diagram of the RN-dS black hole.
  The blue curves are the accretion disks in universe $A$ and $B$, respectively.
  The observer is located near the cosmological horizon.
  The green curve is the ray from the accretion disks in universe $B$ and received by the observer.
  And the red curve is the ray from the accretion disks in universe $A$.}
  \label{penrose}
\end{figure}

The position of the observer will affect the image of the black hole, especially the ray with large $b$ will not be received by the observer.
Therefore, we set the observer far from the black hole and near the cosmological horizon in this section, i.e., $r_\mathrm{obs}=0.9r_\mathrm{c}$.
A part of the Penrose diagram of the RN-dS black hole is given by Fig.\ref{penrose}.
If the Cauchy horizon is stable, the infinite repetition of the spacetime allow the ray emitted from the accretion disk in the previous universe $A$ to fall into the black hole, fly out from the white hole and be received by the observer in universe $B$ finally.
This kind of trajectory of the photon is shown as the red curve in Fig.\ref{penrose}.
Besides, the photons can be emitted from the accretion disk and received by the observer both in universe $B$, as the green curve in Fig.\ref{penrose}.
The images caused by this two kinds of ray are quite different.
This can be found in the following discussions.

\subsection{The appearance of the RN-dS black hole with ISCO and OSCO}
\begin{figure}[htbp]
\centering
\subfigure[]{\includegraphics[width=4cm]{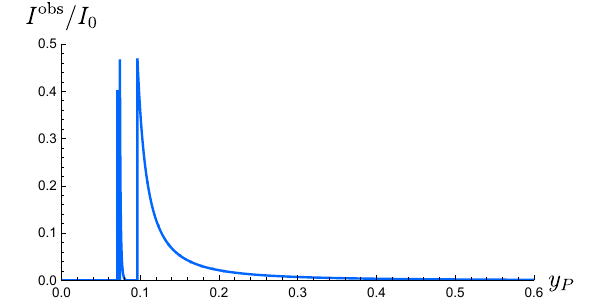}}
\subfigure[]{\includegraphics[width=4cm]{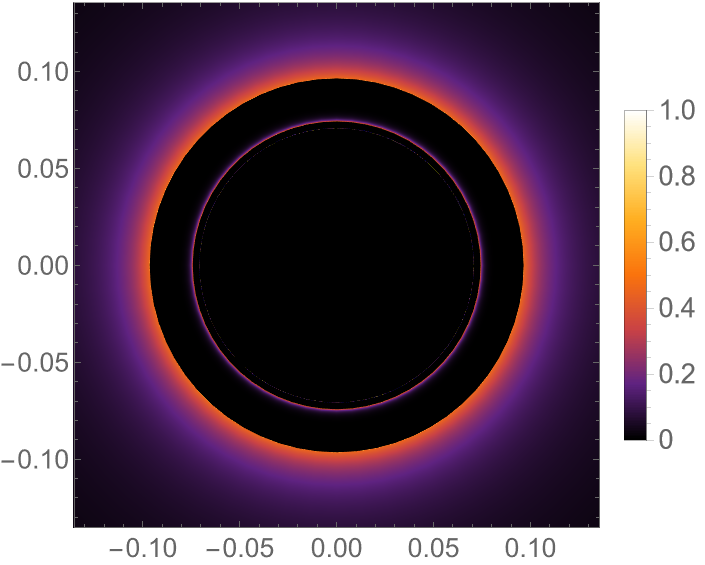}}
\subfigure[]{\includegraphics[width=4cm]{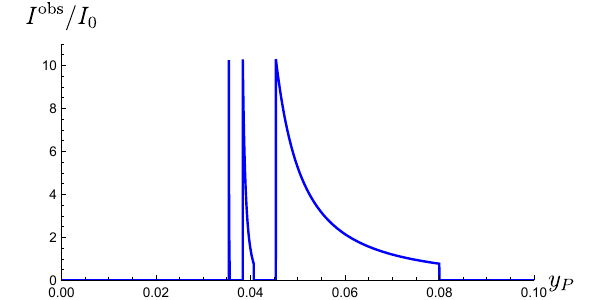}}
\subfigure[]{\includegraphics[width=4cm]{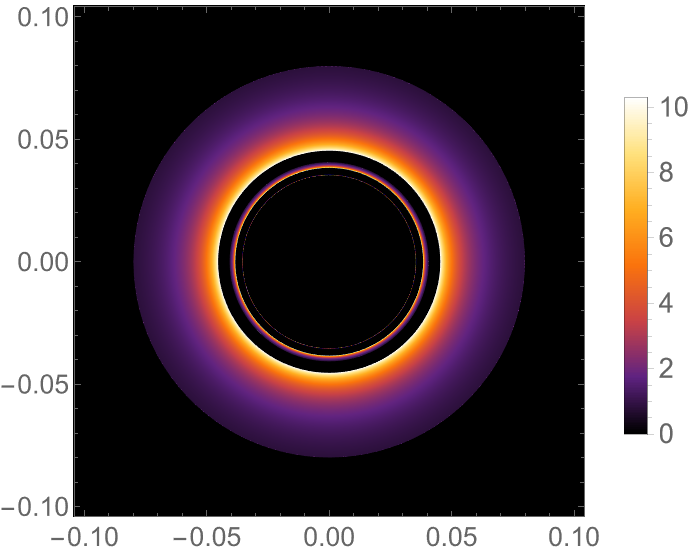}}

\subfigure[]{\includegraphics[width=4cm]{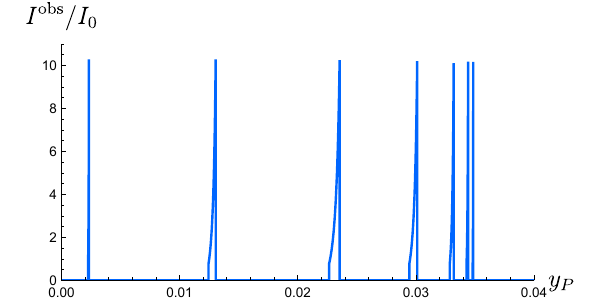}}
\subfigure[]{{\includegraphics[width=4cm]{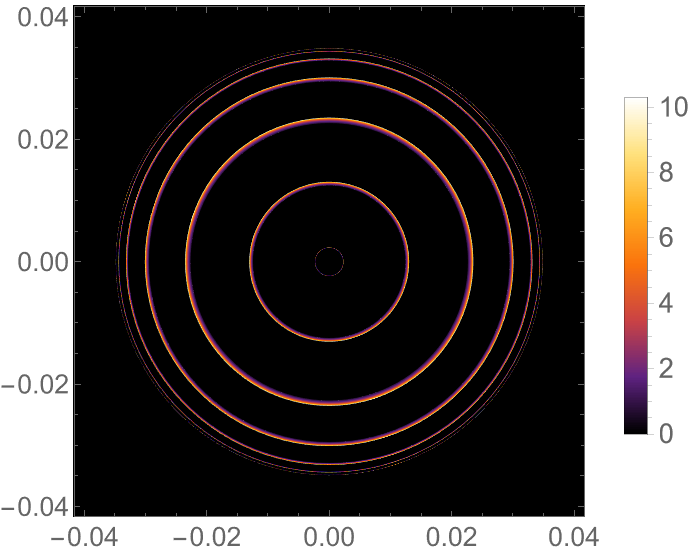}}\label{WHRNdS}}
\subfigure[]{\includegraphics[width=4cm]{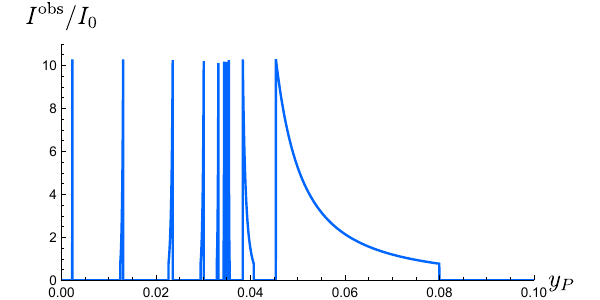}}
\subfigure[]{{\includegraphics[width=4cm]{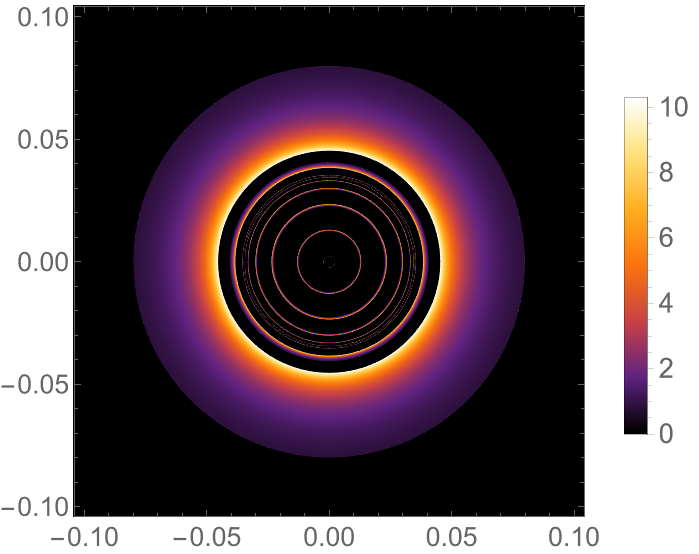}}\label{WHBHRNdS}}
\caption{The intensity and the image of the Schwarzschild black hole with the photon sphere at $r_\mathrm{sp}=1.9996$ ($M=0.6665$) and the RN-dS black hole with $Q=1.0001, \Lambda=0.001$.
Figure (a) and (b) are the observed intensity and the image of the Schwarzschild black hole with photon sphere whose radius is same as that of the RN-dS black hole.
Figure (c) and (d) are the observed intensity and the image of the RN-dS black hole when the photons are emitted from the accretion disk in universe $B$.
Figure (e) and (f) are the observed intensity and the image of the RN-dS black hole when the photons are emitted from the accretion disk in universe $A$.
Figure (g) and (h) are the observed intensity and the image of the RN-dS black hole when the photons are emitted from the accretion disk in both universe $A$ and $B$.}
\label{imageRNdS}
\end{figure}

The appearance of the RN-dS black hole with $Q=1.0001, \Lambda=0.001$ is given by Fig.\ref{imageRNdS} and the observer is located at $r_\mathrm{obs}=48.378$. $r_{\text{obs}}/r_{\text{c}}=0.9$
The light source distributes between ISCO and OSCO and the luminous intensity of the accretion disk is described by Eq.(\ref{IemrISCO}).

Compare to that of the Schwarzschild black hole, the image of the RN-dS black hole has a significant edge.
This is because there is no light source outside the OSCO.
This has already been encountered in Schwarzschild-dS black hole with an ISCO and OSCO.
Besides, the lensed ring or the photon ring of the RN-dS black hole is smaller than that of the Schwarzschild black hole, although their photon spheres are the same size.
The intensity of the Schwarzschild black hole is weaker than the RN-dS black hole, which is caused by redshift factor.
More precisely, when the observer is located far from the black hole, the metric function $f_\mathrm{Sch}(r_\mathrm{obs})$ tends to $1$, while $f_\mathrm{RNdS}(r_\mathrm{obs})$ tends to $0$.
The cosmological constant chosen by us is $\Lambda=0.001$.
This means the magnitude of $\Lambda M^2$ is $10^{-3}$, in other words, it is a supermassive black hole since the astronomical observations have revealed a tiny value for the cosmological constant.
From Eq.(\ref{Iobs}), the intensity is greatly amplified when the observer is very close to the cosmological horizon because of the nearly divergent redshift factor.
Therefore, if the observer is close to the cosmological horizon enough, the intensity emitted near the OSCO is magnified so much that an edge can be seen clearly in the image.
In short, the images of the supermassive RN-dS black holes may have significant edge.

If the accretion disk is located at the universe $A$, then the multi-ring structure occurs.
The rings insides the photon ring are separated obviously and we can distinguish them easily.
The reasons for the formation of the multi-rings structure have been described in detail in \cite{Cao:2023par}.
The key is that the stable Cauchy horizon allows photons to cross it and the infinitely repetitive spacetime allows photons to travel through different universes.
As we had mentioned, this breaks the SCCC.
In some special cases, if the accretion disks are located at both universe $A$ and $B$, then the image is shown in Fig.\ref{WHBHRNdS}.
Many rings occur inside the
area where is traditionally thought to be the shadow, and the closer to the photon ring, the denser the rings become.
And it also has a significant edge.
Therefore, if the SCCC is broken down, then this novel phenomenon may be observed.
However, the multi-ring structure not only occurs when there is a stable Cauchy, but also occurs in case of compact objects and wormholes \cite{Olmo:2021piq,Guerrero:2022qkh,Guerrero:2022msp,Olmo:2023lil}.
In fact, when the effective potential $V(r)$ of the spacetime have a local maximum, the photon sphere (also called critical curve) occurs and there is a bright ring in the image.
Furthermore, if there is an another higher maximum inside the outer one, or $V$ is diverging at somewhere inside the photon sphere, then the multi-ring structure appears \cite{Guerrero:2022msp}.
To distinguish compact objects, wormholes and the black hole with a stable Cauchy horizon, we require additional observation.

\subsection{The appearance of the RN-dS black hole without ISCO and OSCO}
\begin{figure}[htbp]
\centering
\subfigure[]{\includegraphics[width=5cm]{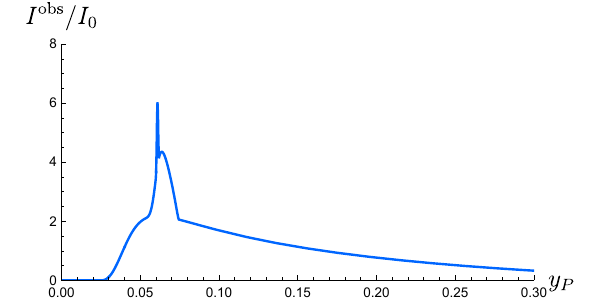}}
\subfigure[]{\includegraphics[width=5cm]{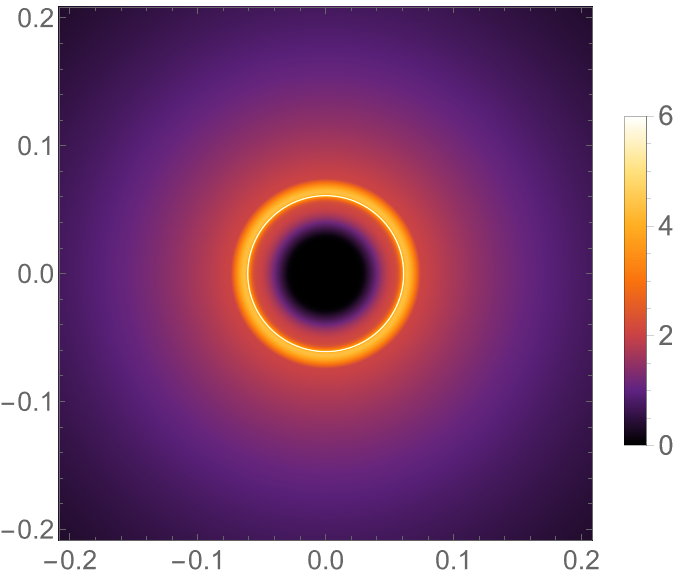}}

\hspace{0.1cm}

\subfigure[]{\includegraphics[width=5cm]{{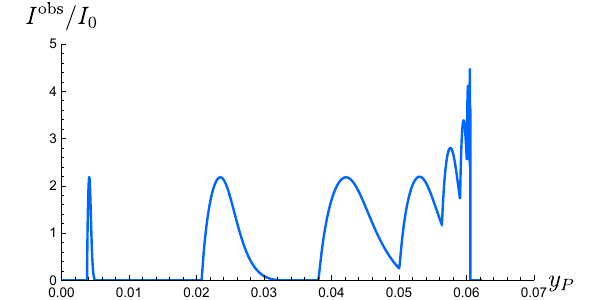}}\label{IWHRNdSrh}}
\subfigure[]{{\includegraphics[width=5cm]{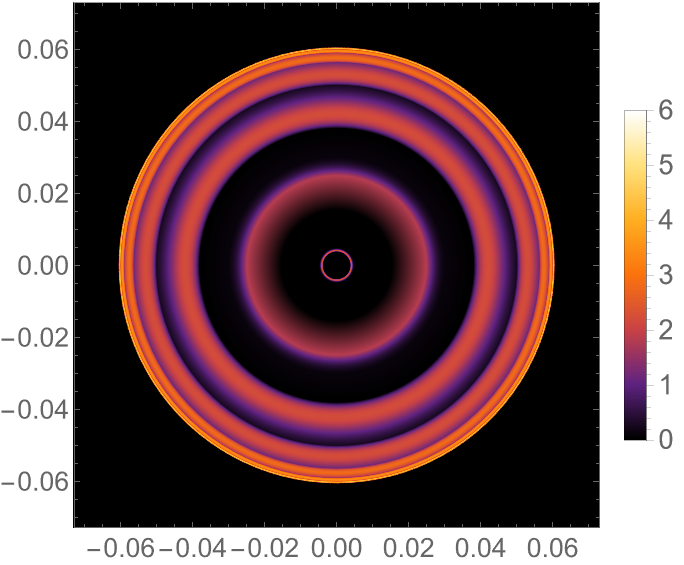}}\label{WHRNdSrh}}

\hspace{0.1cm}

\subfigure[]{\includegraphics[width=5cm]{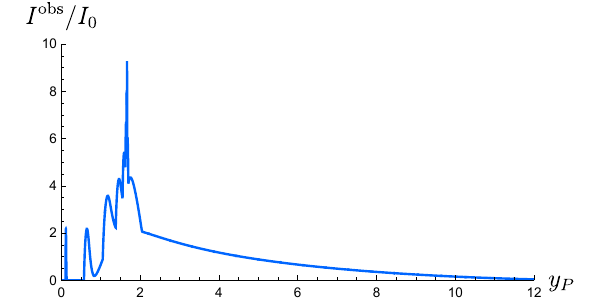}}
\subfigure[]{{\includegraphics[width=5cm]{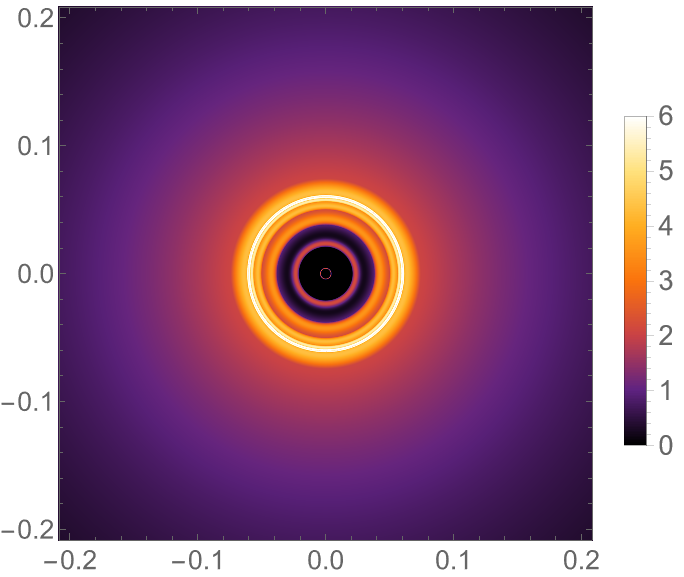}}\label{WHBHRNdSrh}}
\caption{The intensity and the image of the RN-dS black hole with $Q=1.0002, \Lambda=0.003$.
Figure (a) and (b) are the observed intensity and the image of the RN-dS black hole when the photons are emitted from the accretion disk in universe $B$.
Figure (c) and (d) are the observed intensity and the image of the RN-dS black hole when the photons are emitted from the accretion disk in universe $A$.
Figure (e) and (f) are the observed intensity and the image of the RN-dS black hole when the photons are emitted from the accretion disk in both universe $A$ and $B$.}
\label{imageRNdSrh}
\end{figure}

We have already discussed the situation of RN-dS black hole with an ISCO and an OSCO.
However, there are also RN-dS black holes that have no ISCO or OSCO.
The accretion disk around these black holes will have no significant edge.
In this subsection, we will study the image of these black holes.
The luminous intensity of the accretion disk is (\ref{Iemrp}).
The image of the RN-dS black hole with $Q=1.0002, \Lambda=0.003$ is given by Fig.\ref{imageRNdSrh}.
The observer is located at $r_\mathrm{obs}=27.530=0.9r_c$.

\begin{figure}[htb]
  \centering
  \subfigure[]{\includegraphics[width=5cm]{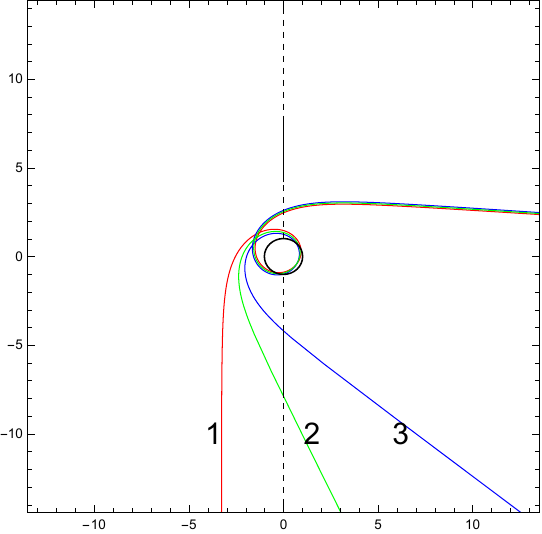}}
\hspace{1cm}
\subfigure[]{\includegraphics[width=5cm]{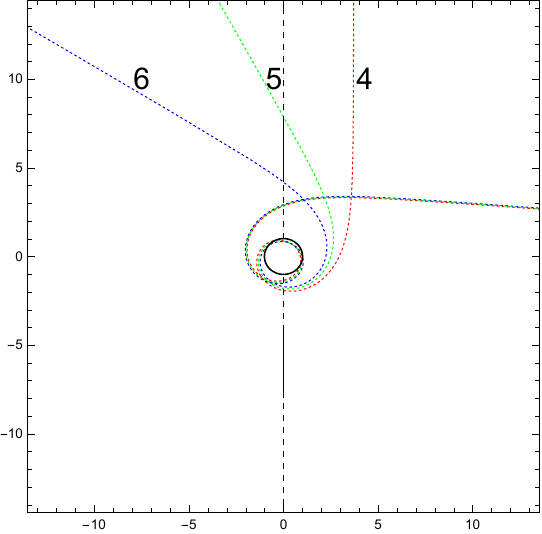}}
  \caption{The trajectory of rays around the RN-dS black hole.
  The black ring is the event horizon.
  The black line is the the accretion disk between ISCO and OSCO, and the dashed black line is the accretion disk between $r_+$ and $r_\mathrm{c}$.
  The rays are for $n=7/4$ (the three solid rays labeled 1$\sim$3 in (a)) and $n=9/4$ (the three dashed rays labeled 4$\sim$6 in (b) )}.
  \label{traj3}
\end{figure}

For the RN-dS black hole without ISCO and OSCO, the shape of the image is similar to that of the Schwarzschild one if the accretion disk is located at universe $B$.
There is no outer edge in the image.
However, for the case of the accretion disk located at universe $A$, or at both $A$ and $B$, the multi-rings structure appears.
Like the image in Fig.\ref{WHBHRNdS}, there are many rings inside the shadow.
And, similarly, the closer to the photon ring (or lensed ring), the denser the rings become.
Compare to Fig.\ref{WHRNdS} and \ref{WHBHRNdS}, the rings in Fig.\ref{WHRNdSrh} and \ref{WHBHRNdSrh} are more wider.
Besides, there is a little bit of overlap between some of the neighboring rings.
In \cite{Guerrero:2022msp}, it had also mentioned this similar phenomenon that, there are wider rings in the image if the edge of accretion disk is closer to the center of object. Here, to explain this phenomenon, the trajectory of rays has been drawn in Fig.\ref{traj3}.

The red, green, and blue solid rays are for the third peak (ring) with the normalized number $n=7/4$ in Fig.\ref{IWHRNdSrh}.
And the red, green, and blue dashed rays are for the fourth one with $n=9/4$.
If the accretion disk is between ISCO and OSCO (black line), then the third ring is from $b_2$ to $b_3$, where $b_i$ is the impact parameter of the ray with corresponding number labeled in Fig. \ref{traj3}.
And the fourth ring is from $b_5$ to $b_6$.
However, if the accretion disk is distribute outside the event horizon (black dashed line), then the third ring begins at $b_1$ and the fourth ring begins at $b_4$.
Therefore, the rings caused by the accretion disk outside all the event horizon is wider than the rings caused by the the accretion disk between ISCO and OSCO.
Take the green dashed ray (labeled 5) for an example, it intersects with the accretion disk below twice and the accretion disk above twice.
It contributes to the third ring and the fourth ring, too.
Thus, all six rays intersect with the accretion disk below for $n=7/4$, and only the red, green, and blue dashed rays intersect with accretion disk above for $n=9/4$.
This means the fourth ring begins when the third ring is not finished.
Therefore, the third and the fourth ring in Fig.\ref{IWHRNdSrh} have a little bit of overlap.

\section{Conclusions and discussion}\label{S5}

SCCC is an important subject in general relativity.
It guarantees the predictability of the gravitation theory.
However, although some potential observations  violating WCCC is proposed in the RN singularity \cite{Wagner:2023axm},  testing SCCC from astronomical observations has been a challenge.
Hence it is significant to study the observational effect of SCCC.
The Cauchy horizon is the boundary of the region where the initial data can predict.
When the Cauchy horizon is stable, one can pass it and arrive at a area that is not determined by the initial data which violates the SCCC.
RN-dS black hole is one of the most famous black hole with a stable Cauchy horizon.

Unlike Schwarzschild black hole and RN black hole, there may exist an ISCO and an OSCO in RN-dS black hole.
The range of parameter when there is an ISCO and an OSCO in the RN-dS black hole with an stable Cauchy horizon is shown in Fig.\ref{range4}.
The accretion disk distributed between the ISCO and OSCO results in significant inner and outer edges in the image.
Taking into account the positive cosmological constant, static observers only exist in the region inside the cosmological horizon.
So we have to place the observer between the event horizon and the cosmological horizon and use stereographic projection to obtain the image.
To analyse the influence of the observed distant on the image, we first draw up the image of Schwarzschild-dS black hole without ISCO and OSCO viewed by observers at different distance.

When the observer is at a finite distance, the observed intensity is truncated, and the corresponding image is clearly bounded.
This effect is particularly noticeable when the observer is close to the black hole, but it becomes progressively less distinguishable as the distance increases.
This is because, for an observer at a finite distance, there is always light
with extremely large impact parameter that cannot be received.
As the observed distance increases, the intensity decreases and then increases, which resulting from that the denominator of the redshift factor in Eq.(\ref{Iobs}) is nearly zero when the observer is too close to the event horizon and the cosmological horizon.
If the observer is far enough away from the black hole,  e.g., $0.9r_c$, the edge may disappear.
We also draw up the image of Schwarzschild-dS black hole with an ISCO and an OSCO in Fig.\ref{imageSdSIO}.
The image has an outer edge, too.
However, the cause of this edge is different.
Besides the influence of the observed distance, the OSCO would result in an outer edge in the image, too.
The reason is that there is no light source outside the OSCO, and this edge always exists no matter how far the observer is.

In order to avoid the influence of the observed distance, we draw up the image of RN-dS black hole with an ISCO, an OSCO and a stable Cauchy horizon viewed by a far away observer in Fig.\ref{imageRNdS}.
The observer is located at universe $B$, while there are two kinds of light sources.
When the photons is emitted from the accretion disk in universe $B$, the image's shape is similar to that of the Schwarzschild one, except that there is an outer edge in the image of RN-dS black hole, which is resulted from the OSCO.
Besides, the photon ring of the RN-dS black hole is smaller than the Schwarzschild one, even though they have the photon spheres of the same size.
And the intensity of the RN-dS one is much larger than the Schwarzschild one, due to the redshift factor.
If the observer is close to the cosmological horizon enough, the redshift factor tends to infinity and the edge of the image is bright enough to be viewed clearly.
However, when the accretion disk is located in universe $A$, the photons can be received by the observer in universe  $B$ by the black-white hole channel.
These photons produce many extra bright rings in the image.
When the accretion disk is located in both universe $A$ and $B$, the image has a outer edge, and the multi-rings structure inside the traditional shadow area.
The image of the RN-dS black hole without ISCO and OSCO is also shown in Fig.\ref{imageRNdSrh}.
The accretion disk is distributed between the event horizon and the cosmological horizon, and sharply peaked at event horizon.
The multi-rings structure also occurs when  there is only one  accretion disk in universe $A$.
As a result, the widely distributed accretion disk leads to the wider rings inside the shadow.
Some of the neighboring rings even have a little overlap.
In conclusion, the image of RN-dS black hole with a stable Cauchy horizon is much different from the Schwarzschild one.

The RN-dS black hole with a stable Cauchy horizon is one of the simplest examples inconsistent with SCCC.
Based on our analysis, the biggest differences between its image and the Schwarzschild's are the outer edge and the multi-rings structure.
Another example which violating SCCC is the regular black hole with a stable Cauchy horizon.
There is also a multi-rings structure in its image \cite{Cao:2023par}.
Therefore, if SCCC is broken down, this multi-rings structure may be observed.
This provides us a new way to test SCCC in astronomical observation.
It is worth noting that such a multi-ring structure can also appear in case of compact objects and wormholes, which have no horizon.
And we need other methods to distinguish them.

Within classical theory, some black holes have a stable Cauchy horizons.
However, certain studies suggest that in such scenarios, quantum effects will play an important role and eventually render these Cauchy horizons unstable \cite{Cai:1995nt,Cai:1998yp,Hollands:2019whz}.
Nevertheless, whether quantum effects will overturn established classical results remains an unresolved issue.
Probing the multi-ring structure could help examine the SCCC, and further assessing the significance of quantum effects.

Our future research will focus on two intriguing directions.
Firstly, we aim to study the images of rotating black holes with stable Cauchy horizons and further explore the presence of multi-rings structures which is similar to those observed in RN-dS images.
Secondly, we plan to investigate other types of black holes that violate the SCCC.
The images of these more complex black holes may reveal new types of structures in black hole images beyond what is currently known.
We hope that these investigations will provide new methods for testing the SCCC in astronomical observations.

\section*{Acknowledgement}

This work was supported in part by the National Natural Science Foundation of China with grants No.12075232 and No.12247103.
It is also supported by the National Key R\&D Program of China Grant No.2022YFC2204603.

\end{document}